\documentclass[11pt,a4paper]{article}
\usepackage[T1]{fontenc}
\usepackage[utf8]{inputenc}
\usepackage{lmodern}
\usepackage[margin=25mm]{geometry}
\usepackage{amsmath,amssymb}
\usepackage{graphicx}
\usepackage{xcolor}
\usepackage{booktabs,tabularx,array}
\usepackage{microtype}
\usepackage{caption}
\usepackage{placeins}
\usepackage{tikz}
\usetikzlibrary{arrows.meta,positioning,calc,shapes.geometric}
\usepackage[authoryear,round]{natbib}
\usepackage[hidelinks]{hyperref}
\newcommand{\estci}[3]{\begin{tabular}{@{}c@{}}$#1$\\[-2pt]{\scriptsize[$#2$, $#3$]}\end{tabular}}
\newcommand{\tabnote}[1]{\par\smallskip\begin{minipage}{\linewidth}\footnotesize #1\end{minipage}}

\title{Religious expression and well-being across countries and regions:\\Evidence from a decade of social media posts}
\author{
Stefano M. Iacus\thanks{Corresponding author: Stefano M. Iacus. Email: \texttt{stefano.iacus@ec.europa.eu}.} \\
   Joint Research Centre \\
  European Commission,
  Seville (SP) \\
  \texttt{stefano.iacus@ec.europa.eu}
  \and
  Giuseppe Porro\\
  Department of Law, Economics and Culture\\
  University of Insubria\,
  Como (IT)\\
  \texttt{giuseppe.porro@uninsubria.it}
  \and
  Haodong Qi \\
  Department of Global Political Studies \\
  Malmö University,
  Malmö (SE)\\
  \texttt{haodong.qi@mau.se}   
}

\begin{document}
\maketitle

\begin{abstract}
This observational study examines whether social connectedness constitutes a pathway between religious expression and subsequent well-being. Drawing on millions of social-media posts, we combine broad international coverage, subnational geographic detail, and annual observations over 2013--2022 to examine religious expression, connectedness, and three well-being dimensions. The longitudinal models account for baseline connectedness and well-being, geographic and year effects, and national trends. National longitudinal models support positive indirect associations through social connectedness for \emph{happiness} and \emph{life satisfaction}. The indirect estimates for \emph{purpose} nationally and for all three outcomes regionally are positive but imprecise. Complementary coefficient contrasts in contemporaneous regional models and comparisons of regional period averages show stronger associations of religious expression with \emph{purpose} than with either \emph{happiness} or \emph{life satisfaction}. Together, these findings support a national statistical pathway through social connectedness for the affective and evaluative dimensions of well-being. The stronger regional associations with \emph{purpose} motivate further investigation of meaning- and direction-related processes beyond the measured connectedness pathway.

\end{abstract}
\noindent\textbf{Keywords:} religious expression; social connectedness;
subjective well-being; \emph{purpose in life}; social media.

\clearpage

\section{Introduction}
\label{sec:introduction}

Religious life connects beliefs and sources of meaning with social relationships, shared practices, and belonging. Its relationship with well-being may therefore operate through different resources rather than through a single, uniform association. Religious-service attendance and congregational friendships have been linked to \emph{life satisfaction} \citep{lim2010religion}, while social support, respect, and meaning have been identified as intermediaries between religiosity and well-being \citep{diener2011religion}. These relationships also depend on social context: national norms concerning religiosity shape the strength of individual-level associations \citep{stavrova2013religious}. Together, these findings point to social resources as one possible pathway linking religious life and well-being, while also cautioning against treating religion as a uniform predictor.

The economic literature reinforces the value of examining social and meaning-related resources alongside material conditions. \citet{CampanteYanagizawaDrott2015} link longer Ramadan fasting to greater subjective well-being among Muslims despite lower output growth in Muslim countries. \citet{Becchetti2025} examines relational and meaning-related channels in European survey data. At the same time, \citet{Prati2024} reports very small or negligible associations for several religious predictors in large survey analyses. Taken together, these studies motivate approaches that identify specific pathways between religious dimensions and well-being outcomes and report the magnitude and uncertainty of the estimated relationships.

Much of this evidence concerns individuals, religious participation, or relationships within congregations. A related question is whether religious expression is associated with subsequent well-being through broader social connectedness at the territorial level. Public expression on social media makes it possible to examine this question across places and periods. \citet{RitterEtAl2014} connect differences in emotional language between Christian and atheist Twitter users partly to language concerning social relationships. Geographic language measures extend such comparisons across places and periods, although their interpretation depends on how indicators are constructed and validated \citep{JaidkaEtAl2020}. Large volumes of text do not eliminate selection into platform use, public expression, or geographic identification. The resulting indicators describe expressions among the observed posting population rather than representative population averages. Their value for the present study lies in allowing the same constructs to be examined repeatedly over time and across geographies, while recognizing the associated selection and measurement limitations.

We use the Human Flourishing Geographic Index (HFGI) framework, which identifies online expressions across dimensions of flourishing \citep{iacus2026hfgi}, to construct country and first-level regional indicators covering 2013--2022. The religious-expression index summarizes expressions about religious beliefs and comfort. The connectedness index combines societal belonging, reversed loneliness, trust in others, and feeling trusted. It concerns the wider social environment rather than relationships specifically formed within religious congregations. The underlying scores reflect both how often each dimension is expressed and the valence of those expressions, rather than population-average well-being. Accordingly, the analysis concerns territorial patterns in online expression, not individual-level effects or representative population averages.

Our primary question is whether social connectedness constitutes a time-ordered indirect statistical pathway between religious expression and later well-being. We distinguish three outcomes: \emph{happiness}, an affect-oriented outcome; \emph{life satisfaction}, which concerns the evaluation of one's life; and \emph{purpose}, which concerns direction and aims. These outcomes reflect the broader distinction among affective, evaluative, and eudaimonic dimensions of well-being \citep{MartelaSteger2016,oecd2025wellbeing}. Social relationships may be relevant to all three outcomes, while religious frameworks of meaning motivate a potentially stronger association with \emph{purpose}. In our specification, \emph{purpose} --- like \emph{happiness} and \emph{life satisfaction} --- is an outcome, and connectedness is the proposed statistical intermediary. This specification complements, but is distinct from, research that places meaning itself between religiousness and well-being \citep{steger2005meaning}.

The analysis begins with contemporaneous national associations (\textbf{H1}) and then examines the connectedness pathway (\textbf{H2}). The longitudinal model relates religious expression in a baseline year to connectedness one year later and well-being two years later, conditional on baseline connectedness and all three outcomes, fixed effects, and national trends. We extend these questions to regions (\textbf{H1-Reg} and \textbf{H2-Reg}) and assess the sensitivity of contemporaneous regional associations to country-by-year effects. A comparison of regional period averages examines whether regions with more positive average religious expression also have higher average expressed well-being than other regions in the same country (\textbf{H3}). This sequence distinguishes annual variation within places from differences between regional averages. Direct contrasts test whether the association with \emph{purpose} exceeds those with the other outcomes (\textbf{H4}).

The national results support positive indirect associations through connectedness for \emph{happiness} and \emph{life satisfaction}; the indirect association for \emph{purpose} and all total lagged associations remain imprecise. The regional lagged pathway estimates are also imprecise. Positive contemporaneous associations appear for all outcomes at both scales, and the regional findings remain under country-by-year controls. Associations between regional averages are positive as well. The direct contrasts support a stronger regional association with \emph{purpose} than with either other outcome; nationally, the estimated contrast with \emph{life satisfaction} remains imprecise.

This study extends research on religion and well-being to repeated territorial patterns of online expression. It combines international and subnational observations over a decade to distinguish time-ordered associations from persistent geographic differences, examine social connectedness as a possible statistical pathway, and compare associations across affective, evaluative, and \emph{purpose}-related dimensions of well-being. The findings suggest that connectedness may contribute to the territorial relationship between religious expression and later \emph{happiness} and \emph{life satisfaction}. The stronger association with \emph{purpose} points to a distinct pattern, while the imprecise indirect estimate leaves open whether other meaning- or direction-related processes account for it.

The remainder of the paper is structured as follows. Section~\ref{sec:literature} reviews the related literature. Section~\ref{sec:methods} describes the data and measures, develops the hypotheses and presents the empirical strategy. Section~\ref{sec:results} reports the results. Section~\ref{sec:limitations} addresses the limitations, and Section~\ref{sec:discussion} discusses the findings and presents the conclusions.

\section{Literature review}
\label{sec:literature}

\subsection{Religiousness and the magnitude of well being associations}

Religiousness encompasses affiliation, belief, subjective importance and participation. These dimensions need not have the same relationship with well-being. \citet{diener2011religion} find that the association is generally small after accounting for life circumstances and varies with societal conditions and religiosity. Social support, respect and meaning help account for the observed relationship. Their findings place context and intermediate resources at the centre of the explanation. They also show why country differences in average religiousness cannot be read straightforwardly as evidence about the benefits of individual religiousness.

Recent evidence strengthens the case for distinguishing statistical detectability from substantive importance. \citet{Prati2024}, using international surveys and longitudinal studies, reports very small or negligible associations for several religious predictors. This challenges an interpretation based mainly on whether coefficients cross a significance threshold. It does not imply that every religious dimension is irrelevant in every context, or that territorial expression must exhibit the same effect sizes as individual reports. For the present study, it motivates reporting magnitudes and uncertainty separately and investigating specific social pathways without presuming that they must produce large overall associations.

\subsection{Social relationships and communal resources}

The social explanation receives direct support from \citet{lim2010religion}. Using a two-wave US panel, they show that close friendships within congregations account for much of the association between religious attendance and \emph{life satisfaction}. Religious identity conditions this relationship: congregational friendships are especially relevant when religion is important to the individual's sense of self. The result connects the structure of social relationships to the meanings attached to them. It does not establish that any increase in contact, irrespective of identity or context, provides an equivalent resource.

The economic literature interprets shared experiences and relationships through the concept of relational goods. \citet{Becchetti2025} studies several possible channels connecting religiousness and \emph{life satisfaction}, including relational interaction and life meaning, and examines regional contextual associations. His repeated survey waves permit comparisons across periods, even if they do not follow the same respondents over time. In a pandemic setting, \citet{RubertoEtAl2025} consider social and family relationships, spirituality, religiosity, and trust in institutions together. Their findings motivate distinguishing religious and spiritual dimensions and institutional trust from interpersonal connectedness.

A related literature considers whether resources buffer hardship. \citet{Walsh2024} finds that social meetings, religiousness and political trust are associated with \emph{life satisfaction} and can moderate financial hardship for some groups. This is a moderation question: the association between hardship and well-being varies with available resources. Our mediation question concerns whether religious expression is associated with later connectedness and whether connectedness is associated with subsequent well-being. The distinction matters because evidence for buffering does not, by itself, identify an indirect path from religious expression to well-being.

We retain the social insight of these studies while broadening its empirical referent. Societal belonging, loneliness and trust can capture relationships beyond religious organizations. They may reflect community resources available to religious and nonreligious residents alike. However, our index cannot establish whether connections originate in religious participation, whether they connect people across religious boundaries, or whether shared religious identity strengthens their value. The proposed mediator is therefore broad social connectedness rather than observed congregational integration.

\subsection{Meaning \emph{purpose} and affective experience}

The existential explanation concerns the interpretive resources provided by religion. \citet{steger2005meaning} find that meaning in life statistically mediates associations between religiousness and well-being in student surveys and daily diaries. This supports attention to meaning as a distinct construct. It differs from our specification, in which \emph{purpose} is an outcome and connectedness is the proposed mediator. A study that places meaning between religion and \emph{life satisfaction} cannot be treated as having tested a pathway from religion through connectedness to \emph{purpose}.

\citet{MartelaSteger2016} distinguish coherence, \emph{purpose} and significance within meaning in life. \emph{Purpose} concerns direction and aims; coherence concerns comprehensibility; significance concerns the perceived value of one's existence. This distinction limits the interpretation of the HFGI \emph{purpose} indicator. It represents a component of eudaimonic well-being rather than a complete measure of meaning or flourishing. The \citet{oecd2025wellbeing} framework similarly distinguishes life evaluation, affect and eudaimonia. Comparing the three outcomes can reveal whether religious expression has a similar association with feeling happy, evaluating life favourably and experiencing direction.

Other pathways also remain plausible. \citet{VanCappellenEtAl2016} identify a role for self-transcendent positive emotions in the relationship between religion or spirituality and well-being. Consequently, an association remaining after adjustment for connectedness cannot be assigned uniquely to existential meaning. Conversely, an imprecise connectedness-mediated path for \emph{purpose} does not establish that social relationships are unimportant for \emph{purpose}. Our comparisons distinguish estimated associations and their precision while leaving unmeasured emotional and existential processes open.

\subsection{Cultural context and responses to adversity}

Community resources operate within wider cultural environments. \citet{stavrova2013religious} show that the association between religiousness and well-being is stronger where religiousness is socially normative; social recognition partly accounts for this pattern. This mechanism differs from simply having more friends or receiving support. It concerns how identities are valued. Together with \citet{diener2011religion}, the study cautions against assuming that a religious expression has the same social implications across societies. Our fixed effects account for stable contextual differences, even if they do not test whether religious norms moderate the expression--well-being association.

Longitudinal research also suggests that the relationship can depend on adversity and the religious dimension measured. \citet{ChenEtAl2020} examine religious-service attendance -- i.e., a specific aspect of religiosity -- and subsequent health and well-being outcomes across three prospective cohorts. \citet{IyerEtAl2025} find that pre-pandemic religious belonging and the personal relevance of religion are associated with better coping during UK lockdowns, while greater religious intensity does not confer a uniform additional advantage. These studies support attention to temporal ordering and baseline conditions. They do not imply that the same relationships will appear in aggregate online expression or that findings about psychological distress transfer directly to positive well-being.

\citet{CampanteYanagizawaDrott2015} provide a different economic perspective by exploiting variation in Ramadan fasting duration. Their findings connect more demanding practice to greater subjective well-being despite economic costs. This evidence illustrates why welfare cannot be inferred from material outcomes alone and why participation may have value within a community. Its focus on religious practice complements our analysis of territorial religious expression and motivates attention to social and existential resources.

\subsection{Online expression and geographic well being}

\citet{RitterEtAl2014} provide an early direct antecedent for studying religion and \emph{happiness} on Twitter. Their comparison of Christian and atheist users links differences in emotional language partly to language concerning social relationships and thinking style. Religious identity is inferred through a specific user-selection strategy, and \emph{happiness} is assessed through expression. The study establishes that a social pathway in online religious differences is already part of the literature. Our question extends this approach to repeated territorial indicators and to \emph{life satisfaction} and \emph{purpose} alongside \emph{happiness}.

Geographic language models require a separate measurement argument. \citet{JaidkaEtAl2020} demonstrate that dictionary-based and data-driven methods can produce markedly different geographic well-being estimates when compared with surveys. Large volumes of text do not remove cultural variation in language or sampling bias. We therefore interpret our indicators as measured expressions in the observed posting population. Their correspondence with representative population experience, and their comparability across places and periods, require validation.

The HFGI framework supplies a set of indicators designed to represent expressions relevant to flourishing. The published data descriptor concerns US counties \citep{iacus2026hfgi}. Applying this framework internationally raises further questions about coverage and comparability: using the same labels across languages does not establish measurement equivalence. The common platform and modelling pipeline may also induce shared measurement variation across religious expression, connectedness and well-being. These possibilities remain alternative explanations alongside substantive relationships.

\subsection{Temporal scale and the interpretation of mediation}

Temporal and geographic comparisons answer different questions. A region that has persistently high religious expression may differ from neighbouring regions in institutions, composition or economic conditions. \citet{hamaker2015critique} demonstrate the consequences of conflating stable differences between units with processes within units in longitudinal models.  Their argument motivates separating our annual analyses from comparisons of regional means.

Mediation analysis further separates total, direct and indirect associations. A statistically detectable indirect path does not require a statistically significant total association \citep{ZhaoEtAl2010}. Introducing connectedness can therefore clarify how an imprecisely estimated total association is decomposed. The size and precision of the total association remain separate empirical questions. Likewise, evidence for an indirect path in one outcome and an interval including zero in another is insufficient to demonstrate a difference between the paths. Such comparisons require a direct contrast that accounts for their dependence.

These considerations define the scope of the present analysis. We examine whether a time-ordered connectedness pathway is detectable at country and regional levels, compare the three well-being outcomes and separately describe persistent differences between regions within countries. Baseline adjustment, fixed effects and trends reduce some alternative explanations, but changing social conditions, platform composition, reverse relationships and measurement error can remain. The resulting decomposition is an exploratory account of territorial associations whose substantive interpretation depends on outcome, temporal design and geographic scale.

\section{Data and methods}
\label{sec:methods}

\begin{figure}[htbp]
\centering
\includegraphics[width=\linewidth]{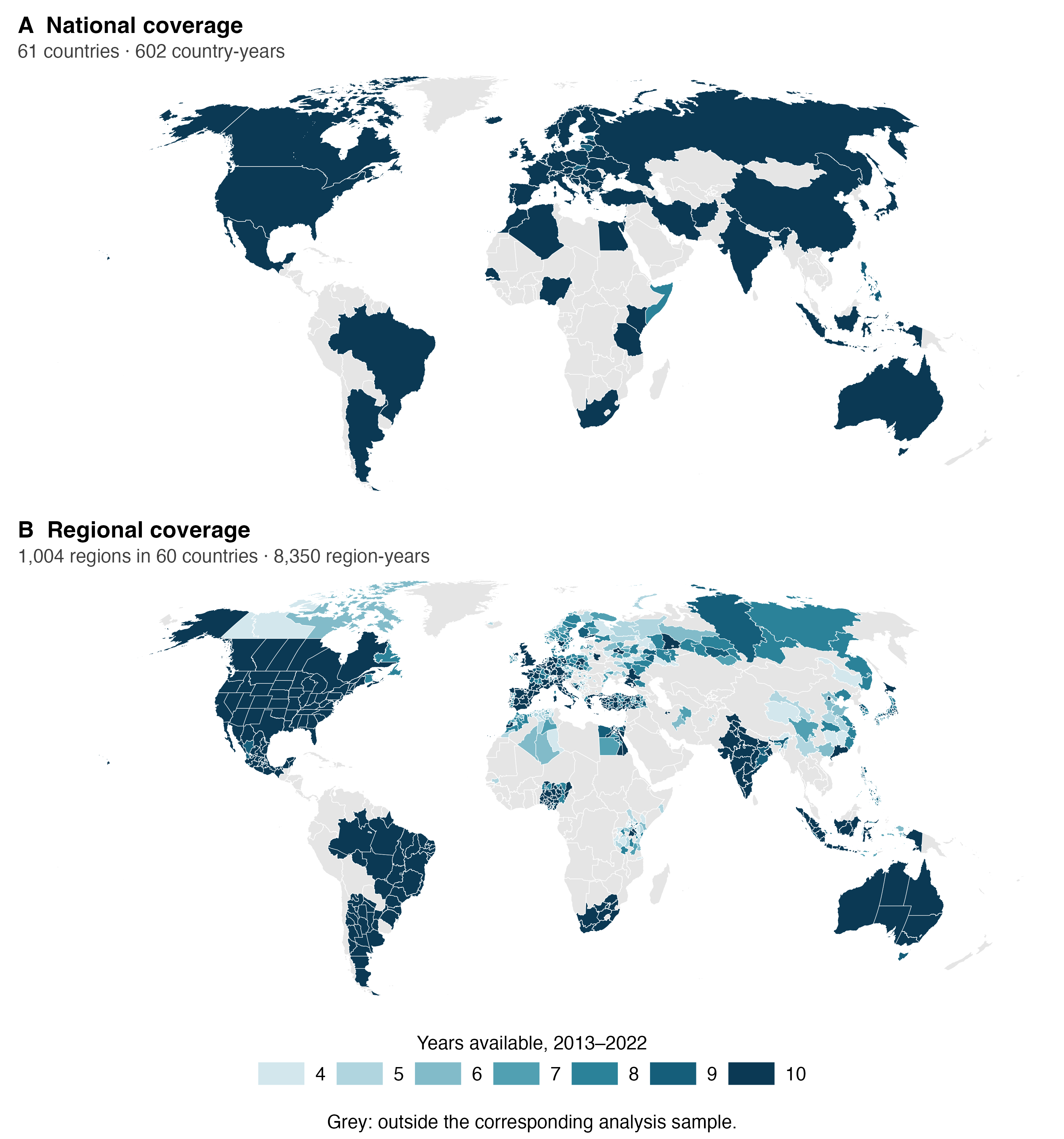}
\caption{Geographic coverage of the contemporaneous association samples, 2013--2022. Panels A and B show the national and regional samples, respectively. Colours indicate the number of eligible annual observations for each geographic unit; grey areas are outside the corresponding sample. The longitudinal and regional period-average analyses apply additional inclusion criteria and use different samples, as detailed in Supplementary Section~\ref{app:data-preparation}.}
\label{fig:coverage}
\end{figure}

We use annual country and regional indicators from the Human Flourishing Geographic Index (HFGI), derived from geographically located Twitter posts. The HFGI framework identifies expressions relating to different dimensions of flourishing and summarizes their positive or negative valence. Its construction is described by \citet{iacus2026hfgi}. Our analyses use these aggregate indicators for 2013--2022. Regions correspond to first-level administrative divisions within countries, whose size and institutional role vary across national settings. The United States enters as one country, with the 50 states and the District of Columbia as regional units. Figure~\ref{fig:coverage} shows the geographic coverage and the number of annual observations available for the contemporaneous association analyses.

The measures describe how people express experiences and beliefs in online posts. For each dimension, we sum the scores assigned to posts in which it is expressed and divide by the total number of posts in the geographic unit and year. Each indicator therefore combines the share of posts expressing that dimension with the average valence of those expressions. Religious expression in this study refers to this combined measure, rather than to the proportion of residents who are religious. Similarly, the well-being indicators describe expressed experiences rather than population averages obtained through a representative survey.

We examine three \emph{well-being} outcomes: \emph{happiness} (\texttt{happiness}), \emph{life satisfaction} (\texttt{lifesat}), and \emph{purpose in life} (\texttt{purpose}). These distinguish expressions of feeling happy, evaluating one's life, and experiencing direction or aims. The \emph{religious-expression} index combines four dimensions: \emph{belief in an afterlife} (\texttt{afterlife}), \emph{belief in God} (\texttt{believegod}), \emph{feeling loved by God} (\texttt{lovedgod}), and \emph{comfort derived from religion} (\texttt{relcomfort}). Together, they capture expressions concerning \emph{religious beliefs} and the personal reassurance associated with them.  

\emph{Social connectedness} combines \emph{belonging} (\texttt{belonging}), \emph{loneliness} (\texttt{loneliness}), \emph{trust in others} (\texttt{trust}), and \emph{feeling trusted by others} (\texttt{trusted}). Loneliness is reversed so that higher values consistently indicate greater connectedness. Belonging refers to society broadly, rather than membership of a religious congregation \citep{iacus2026hfgi}. This index allows us to examine whether religious expression is associated with subsequent expressions of social connection and whether these, in turn, are associated with later well-being. The indices and outcomes are standardized for analysis; the details are provided in Supplementary Section~\ref{app:data-preparation}.
The implications of selection into online expression and of measurement comparability are discussed in Section~\ref{sec:limitations}.

By applying different models, we examine whether \emph{positive religious expression} is associated with \emph{higher well-being} and whether social connectedness provides an \emph{indirect pathway} linking religious expression to subsequent well-being. Our analysis begins at the country level and then extends to regions within countries. The extended regional analysis allows us to examine whether regions with more positive average religious expression exhibit higher average expressed well-being than other regions within the same country. Because religious beliefs can provide a sense of meaning and direction,
we further test whether religious expression is more strongly associated
with \emph{purpose} than with \emph{happiness} and
\emph{life satisfaction}. 

The following sections formulate these hypotheses and present the models used to assess them. A complementary analysis examines whether differences in average
connectedness help account for the association between average religious expression and average well-being across regions within the same country. Unlike the regional longitudinal analysis, this comparison uses regional averages over 2013--2022.  

\subsection{Do religious expression and well-being vary together within countries?}

Our first hypothesis concerns whether religious expression and well-being vary together over time within countries: 

\medskip
\noindent
\textbf{H1:} \emph{More positive religious expression is associated with higher expressed happiness, life satisfaction, and purpose within countries.} 

\medskip
To assess this hypothesis, we estimate contemporaneous associations while accounting for stable national characteristics, common annual changes, and country-specific trends. The sample contains 602 country-years in 61 countries over 2013--2022; each observation represents one country in one year.
Let $i$ index countries, $t$ years, and $k$ the well-being outcome, with $k\in\{\text{\texttt{happiness}}, \text{\texttt{lifesat}}, \text{\texttt{purpose}}\}$. We denote the religious-expression index by $R_{it}$ and outcome $k$ by $Y^{(k)}_{it}$. In the model below, $\alpha_i$ is a country fixed effect, accounting for stable differences between countries; $\lambda_t$ is a year fixed effect, accounting for changes shared across countries; and $\delta_i$ is the coefficient of a country-specific linear time trend. The coefficient of interest, $\beta_k$, measures the association between religious expression and outcome $k$ after these adjustments, and $\varepsilon^{(k)}_{it}$ is the unexplained component of that outcome:

\begin{equation}
Y^{(k)}_{it}=\alpha_i+\lambda_t+\delta_i t+\beta_k R_{it}+\varepsilon^{(k)}_{it}.
\label{eq:country}
\end{equation}

In these association models, fixed effects and trend coefficients are estimated separately for each outcome; their outcome index is omitted for readability. We give each observation equal weight and standardize the predictor and outcome within the estimation sample. Standard errors are clustered by country, allowing observations from the same country to be statistically dependent. Positive estimates of $\beta_k$ support \textbf{H1} for the corresponding outcome; we report their magnitudes and 95\% confidence intervals (CIs). Each coefficient $\beta_k$ describes the association between
within-country annual deviations in religious expression and outcome $k$,
after adjustment for country and year effects and country-specific
linear trends.
These adjustments also absorb time-invariant differences in religious affiliation (Supplementary Section~\ref{app:affiliation-adjustment}).

\subsection{Does social connectedness link religious expression to subsequent well-being within countries?}

The central question is whether social connectedness helps account for a relationship between religious expression and subsequent well-being:

\medskip
\noindent
\textbf{H2:} \emph{Religious expression has a positive indirect association with subsequent happiness, life satisfaction, and purpose through social connectedness at country level.} 

\medskip
We assess this hypothesis using a longitudinal structural equation model (SEM), which estimates the connectedness and well-being equations jointly. The sample comprises 468 country-years in 61 countries, each linking a baseline year to observations one and two years later within 2013--2022. The smaller sample reflects the need for complete measures across three consecutive years. Inclusion criteria for all models are detailed in Supplementary Section~\ref{app:data-preparation}.

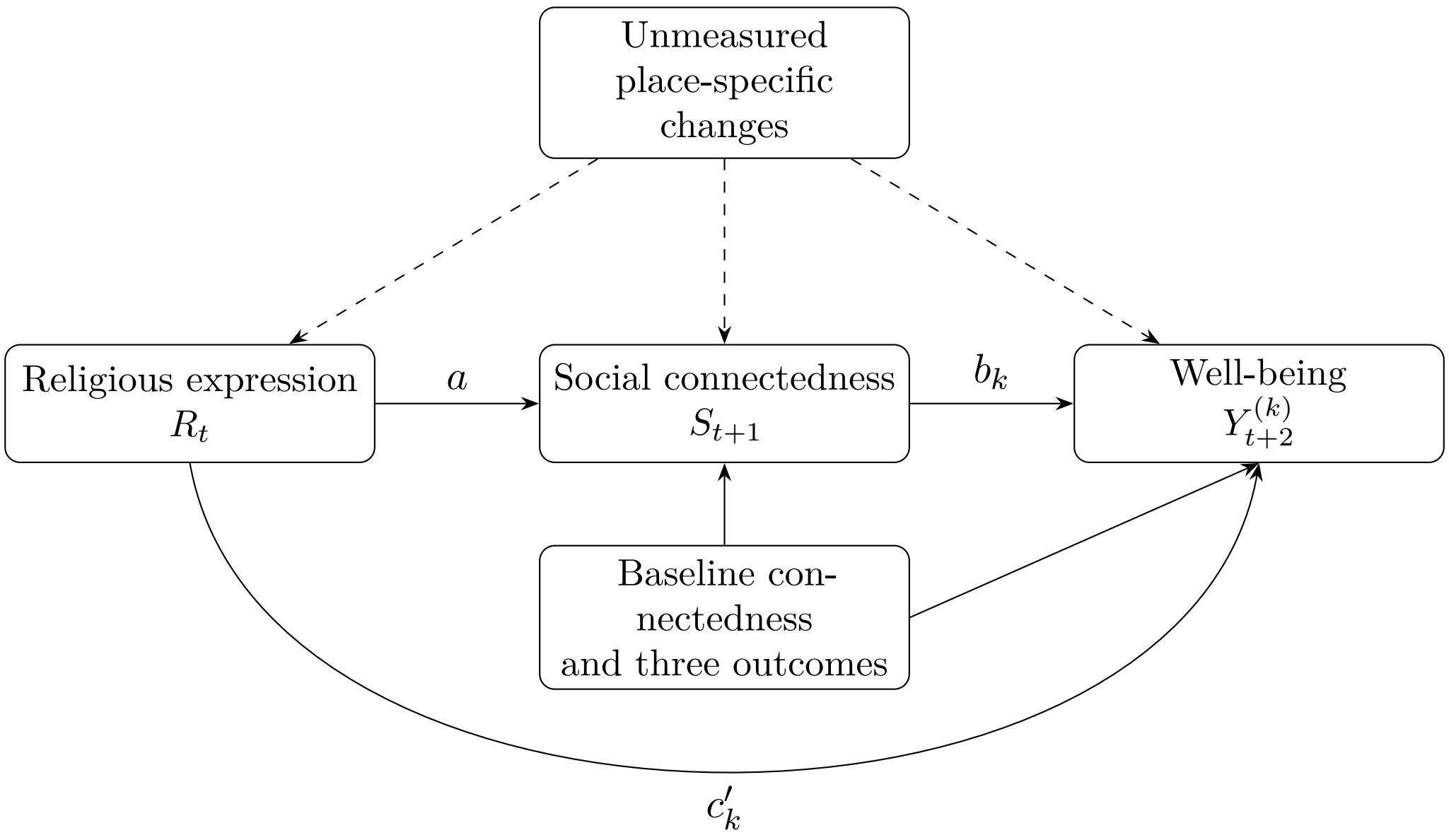
\begin{figure}[htbp]
\centering
\begin{tikzpicture}[>=Stealth, node distance=18mm,
 box/.style={draw,rounded corners,align=center,font=\small,text width=32mm,minimum height=11mm}]
\node[box] (r) at (0,0) {Religious expression\\$R_t$};
\node[box] (s) at (5,0) {Social connectedness\\$S_{t+1}$};
\node[box] (y) at (10,0) {Well-being\\$Y^{(k)}_{t+2}$};
\node[box] (u) at (5,3) {Unmeasured\\place-specific changes};
\node[box] (b) at (5,-2) {Baseline connectedness\\and three outcomes};
\draw[->] (r)--node[above]{$a$}(s);
\draw[->] (s)--node[above]{$b_k$}(y);
\draw[->] (r.south) to[bend right=80] node[below]{$c'_k$}(y.south);
\draw[->,dashed] (u)--(r);
\draw[->,dashed] (u)--(s);
\draw[->,dashed] (u)--(y);
\draw[->] (b)--(s);
\draw[->] (b.east)--(y.south);
\end{tikzpicture}
\caption{Proposed annual pathway at country and regional scales. 
Fixed effects and trends
are omitted from the drawing for clarity.}
\label{fig:dag}
\end{figure}

Denoting the connectedness index by $S$, the proposed sequence is $R_t\rightarrow S_{t+1}\rightarrow Y^{(k)}_{t+2}$: religious expression in year $t$, connectedness one year later, and well-being two years later. We also estimate a religion--outcome path conditional on connectedness. Figure~\ref{fig:dag} illustrates these relationships and additional contextual influences. The coefficient $a$ links religious expression to subsequent connectedness, $b_k$ links connectedness to subsequent outcome $k$, and $c'_k$ represents the remaining religion--outcome association. 

To focus on variation within countries, we first regress each modeled
variable on country and year fixed effects and country-specific linear
trends, using weights that give each country equal total weight.
We then use the residuals in the SEM. These residuals represent the
variation remaining after accounting for the specified effects and
trends, and are denoted by tildes.
We use $g$ to identify the geographic unit: a country here and a region
in the subsequent regional analysis. The vector $B_{gt}$ contains
connectedness and all three well-being outcomes measured in the
baseline year $t$. Including these measures allows us to examine
subsequent associations while accounting for initial connectedness
and well-being.
The vectors $\gamma_S$ and $\gamma_{Y,k}$ contain the coefficients
of these baseline measures in the connectedness and outcome equations,
respectively. The superscript $\top$ denotes transposition, so that
each product with $B$ sums the baseline measures multiplied by their
coefficients. Finally, $e^S_{gt}$ and $e^{Y,k}_{gt}$ denote the
unexplained components of the two equations:
\begin{align}
\widetilde S_{g,t+1} &= a\widetilde R_{gt}
 +\gamma_S^\top\widetilde B_{gt}+e^S_{gt},\label{eq:sems}\\
\widetilde Y^{(k)}_{g,t+2} &= b_k\widetilde S_{g,t+1}
 +c'_k\widetilde R_{gt}+\gamma_{Y,k}^\top\widetilde B_{gt}
 +e^{Y,k}_{gt},\label{eq:semy}
\end{align}

Countries receive equal total weight in the lagged sample. We estimate the equations with robust maximum likelihood with standard errors clustered by country \citep{rosseel2012lavaan}. The residuals of the three outcome equations may covary, allowing unexplained variation in the well-being outcomes to be associated. The model uses the observed indices described above rather than estimating latent factors.

The indirect association relevant to \textbf{H2} is the product $a b_k$. The total modeled association is $c_k=c'_k+a b_k$, combining the conditional religion--outcome coefficient and the indirect component. We report both components and the total, with their uncertainty. The total refers to the lagged sample and is distinct from the same-year coefficient $\beta_k$ used to assess \textbf{H1}. An indirect association can be estimated even when the total association is imprecise \citep{ZhaoEtAl2010}. 
We estimate an explicit mean structure with the four endogenous intercepts fixed at zero after within-country centering. The 95\% confidence intervals use robust standard errors and a normal approximation, with the delta method for indirect and total associations.

\subsection{Do religious expression and well-being vary together within regions?}

National averages can conceal substantial differences between regions
within the same country. Regional analysis allows us to examine whether
the relationships observed nationally also appear in changes within
smaller geographic units. It also allows us to account for annual
changes shared across regions of the same country through
country-by-year effects. The regional extension therefore examines
both the geographic scope of the findings and their sensitivity to
shared national changes. We first consider contemporaneous associations:

\medskip
\noindent
\textbf{H1-Reg:} \emph{More positive religious expression is associated with higher expressed happiness, life satisfaction, and purpose within regions.} 

\medskip
This extends \textbf{H1} to a finer geographic scale. The sample
contains 8,350 region-years in 1,004 regions and 60 countries over
2013--2022. Let $r$ index regions and $i(r)$ identify the country
containing region $r$. Replacing the country fixed effect with
a region fixed effect, $\alpha_r$, gives:

\begin{equation}
Y^{(k)}_{rt}=\alpha_r+\lambda_t+\delta_{i(r)}t+\beta_k R_{rt}+\varepsilon^{(k)}_{rt},
\label{eq:region}
\end{equation}


The outcome and religious-expression index now refer to region $r$
in year $t$. The region fixed effects $\alpha_r$ account for persistent
differences between regions, while $\lambda_t$ accounts for common
annual changes. The coefficient $\delta_{i(r)}$ represents the linear
time trend for the country containing region $r$. It has the same
interpretation as $\delta_i$ in Equation~\eqref{eq:country}, but is
estimated separately in the regional model and need not have the
same value. The coefficient $\beta_k$ measures the association
between religious expression and outcome $k$ within regions,
conditional on these effects and trends. Coefficients are estimated separately in each model; reusing $\beta_k$ does not constrain the national and regional associations to be equal.

To assess whether the regional association persists under a more flexible adjustment for national changes, we replace common year effects and national linear trends with country-by-year fixed effects. Each effect, $\lambda_{i(r),t}$, accounts for changes shared by the observed regions of one country in a particular year. We denote the resulting religious-expression coefficient by $\beta^{\mathrm{CY}}_k$, where CY indicates this country-by-year adjustment:

\begin{equation}
Y^{(k)}_{rt}=\alpha_r+\lambda_{i(r),t}+\beta^{\mathrm{CY}}_k R_{rt}+\varepsilon^{(k)}_{rt}.
\label{eq:region-country-year}
\end{equation}

This comparison uses the same 8,350 observations, weights, and standardization as the preceding regional model, so changes in the coefficients reflect the specification rather than changes in the sample. Standard errors remain clustered by country, using the same small-sample adjustment.
We also repeat the outcome comparisons described below under this specification.

\subsection{Does the connectedness pathway emerge within regions?}

The second regional analysis extends the country-level mediation analysis:

\medskip
\noindent
\textbf{H2-Reg:} \emph{Religious expression has a positive indirect association with subsequent happiness, life satisfaction, and purpose through social connectedness at regional level.} 

\medskip
We use the same temporal ordering and baseline controls as for \textbf{H2}, applying Equations~\eqref{eq:sems} and \eqref{eq:semy} with $g$ now denoting a region. The sample contains 5,122 region-years in 847 regions and 60 countries, with each baseline observation linked to the next two years within 2013--2022.

Before estimating the equations, we remove region and year fixed effects and country-specific linear trends from each variable. Countries receive equal total weight across their regional observations, and standard errors are clustered by country. We report the indirect products, conditional religion--outcome coefficients, and total associations using the same definitions as in the national SEM. This extension examines whether the proposed pathway is also detectable at regional scale. Differences in statistical significance across scales do not themselves establish that the national and regional coefficients differ.

\subsection{Do regions with more positive average religious expression express higher average well-being?}

Annual models examine whether changes in religious expression within
a region accompany changes in well-being. They do not establish whether
regions with higher average religious expression also have higher
average well-being. To address this complementary question, we compare
regional averages over 2013--2022 within each country. Averaging over
the period reduces the influence of short-term fluctuations and allows
us to examine whether the positive association also characterizes
longer-term differences between regions:

\medskip
\noindent
\textbf{H3:} \emph{Regions with more positive average religious expression also exhibit higher average expressed happiness, life satisfaction, and purpose than other regions within the same country.}

\medskip
To assess \textbf{H3}, we relate regional averages of each well-being outcome to average religious expression, accounting for country differences. We require at least eight complete annual observations per region and at least two eligible regions per country. This yields 733 regions in 43 countries. Each year receives equal weight in the period mean; details of index construction are in Supplementary Section~\ref{app:data-preparation}.

For region $r$ in country $i=i(r)$, bars denote period averages and $\alpha_i$ is a country fixed effect. The coefficient $\beta_k$ now describes the association between regional averages of religious expression and outcome $k$, while $\varepsilon^{(k)}_{ri}$ is the unexplained component:

\begin{equation}
\overline Y^{(k)}_{ri}=\alpha_i+\beta_k\overline R_{ri}+\varepsilon^{(k)}_{ri}.
\label{eq:static}
\end{equation}

We give each country equal total weight and cluster standard errors by country. The country fixed-effect terms ensure that the comparison is between regions within the same country. The estimates of $\beta_k$ and their confidence intervals assess the direction, magnitude, and precision of the associations proposed in \textbf{H3}. As a complementary analysis, we examine how connectedness contributes to these associations. We remove country means using the same weights, with tildes denoting the remaining differences. The static observed-variable SEM is:

\begin{align}
\widetilde{\overline S}_{ri} &= a\widetilde{\overline R}_{ri}+e^S_{ri},\label{eq:static-sems}\\
\widetilde{\overline Y}^{(k)}_{ri} &= b_k\widetilde{\overline S}_{ri}
 +c'_k\widetilde{\overline R}_{ri}+e^{Y,k}_{ri}.\label{eq:static-semy}
\end{align}

The coefficients $a$, $b_k$, and $c'_k$ have the same decomposition roles as in the longitudinal model, but here all measures summarize the same period and coefficients are estimated separately. The terms $e^S_{ri}$ and $e^{Y,k}_{ri}$ are the equation residuals, and the outcome residuals may covary. As in the longitudinal SEMs, the intercepts in the connectedness
equation and the three well-being equations are fixed at zero,
because removing country means has already centered each variable. 
The indirect product is $a b_k$, and the total association is $c_k=c'_k+a b_k$. The total association $c_k=c'_k+a b_k$ has the same point estimate
as the coefficient $\beta_k$ in Equation~\eqref{eq:static}.
The regression estimates this association directly, whereas the SEM
decomposes it into a coefficient conditional on connectedness and an
indirect component through connectedness. The confidence intervals
may differ: the regression uses the standard error of $\beta_k$,
while the SEM uses the delta method to calculate uncertainty for
the sum $c'_k+a b_k$. 
This decomposition describes associations among regional averages and complements the test of \textbf{H3}; it does not repeat the temporal tests of \textbf{H2} and \textbf{H2-Reg}.

\subsection{Is religious expression more strongly associated with \emph{purpose} than with \emph{happiness} and \emph{life satisfaction}?}

The account of religion as a source of meaning and direction discussed above motivates a further expectation: its association with \emph{purpose} may be stronger than its associations with feeling happy or evaluating life favourably. We assess this expectation by directly comparing the associations across outcomes:

\medskip
\noindent
\textbf{H4:} \emph{Religious expression is more strongly associated with expressed purpose than with either expressed happiness or expressed life satisfaction.}

\medskip
To assess \textbf{H4}, we compare the \emph{purpose} coefficient with each of the other two outcome coefficients separately in the national annual, regional annual, and regional period-average models. For each comparison, we regress the difference between the two standardized outcomes on religious expression, using a common sample and the same controls and weights as the corresponding association model. A positive coefficient for \emph{purpose} minus another outcome indicates a larger \emph{purpose} association. 
These comparisons test whether the religious-expression coefficient
$\beta_k$ is larger for \emph{purpose} than for \emph{happiness}
or \emph{life satisfaction} in the association models
(Equations~\eqref{eq:country}, \eqref{eq:region}, and \eqref{eq:static}).
They do not test differences between the indirect associations
$a b_k$ through social connectedness estimated by the SEMs.
These six exploratory comparisons use country-clustered 95\% confidence intervals without adjustment for multiple testing.

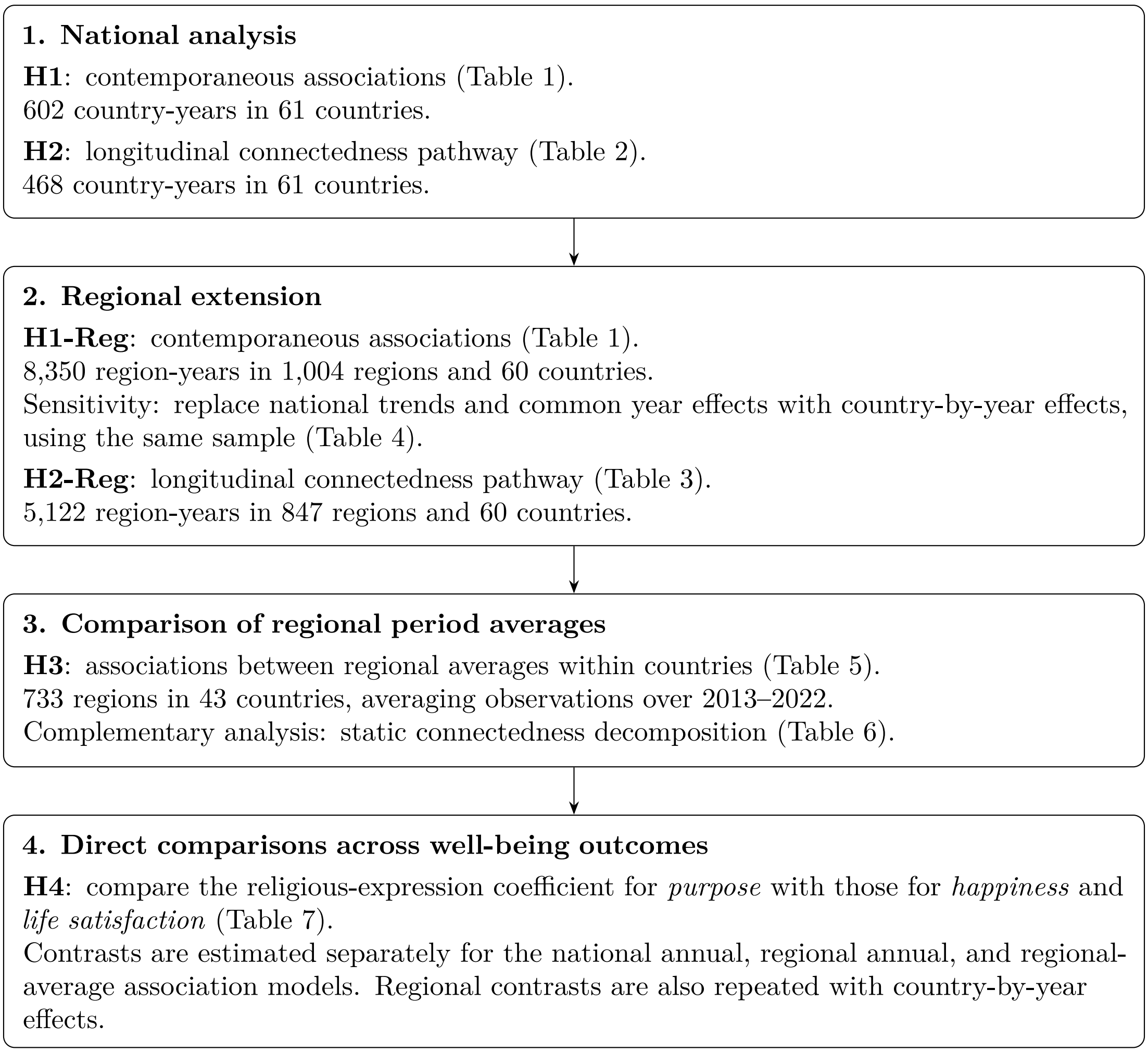
\begin{figure}[htbp]
\centering
\begin{tikzpicture}[
  stage/.style={
    draw,
    rounded corners,
    align=left,
    text width=14cm,
    inner sep=7pt,
    font=\small
  },
  >=Stealth
]

\node[stage] (national) {
  \textbf{1. National analysis}\\[3pt]
  \textbf{H1}: contemporaneous associations
  (Table~\ref{tab:annual}).\\
  602 country-years in 61 countries.\\[3pt]
  \textbf{H2}: longitudinal connectedness pathway
  (Table~\ref{tab:country-sem}).\\
  468 country-years in 61 countries.
};

\node[stage, below=6mm of national] (regional) {
  \textbf{2. Regional extension}\\[3pt]
  \textbf{H1-Reg}: contemporaneous associations
  (Table~\ref{tab:annual}).\\
  8,350 region-years in 1,004 regions and 60 countries.\\
  Sensitivity: replace national trends and common year effects with
  country-by-year effects, using the same sample
  (Table~\ref{tab:country-year-robustness}).\\[3pt]
  \textbf{H2-Reg}: longitudinal connectedness pathway
  (Table~\ref{tab:region-sem}).\\
  5,122 region-years in 847 regions and 60 countries.
};

\node[stage, below=6mm of regional] (averages) {
  \textbf{3. Comparison of regional period averages}\\[3pt]
  \textbf{H3}: associations between regional averages within countries
  (Table~\ref{tab:static}).\\
  733 regions in 43 countries, averaging observations over 2013--2022.\\
  Complementary analysis: static connectedness decomposition
  (Table~\ref{tab:static-sem}).
};

\node[stage, below=6mm of averages] (contrasts) {
  \textbf{4. Direct comparisons across well-being outcomes}\\[3pt]
  \textbf{H4}: compare the religious-expression coefficient for
  \emph{purpose} with those for \emph{happiness} and
  \emph{life satisfaction}
  (Table~\ref{tab:contrasts}).\\
  Contrasts are estimated separately for the national annual,
  regional annual, and regional-average association models.
  Regional contrasts are also repeated with country-by-year effects.
};

\draw[->] (national) -- (regional);
\draw[->] (regional) -- (averages);
\draw[->] (averages) -- (contrasts);

\end{tikzpicture}
\caption{Overview of the estimation strategy.
Arrows indicate the order of presentation.
All analyses use observations from 2013--2022.
The models are estimated separately, with sample sizes determined
by their respective inclusion criteria.}
\label{fig:estimation-strategy}
\end{figure}

Figure~\ref{fig:estimation-strategy} summarizes the estimation strategy.
We first examine national associations and the connectedness pathway,
then extend these questions to regions. The country-by-year specification
checks whether the contemporaneous regional associations persist after
accounting for annual changes shared within countries. We subsequently
compare regional period averages. Finally, we test whether the
religious-expression association is stronger for \emph{purpose} than
for \emph{happiness} and \emph{life satisfaction}. These final contrasts
compare coefficients from the association models, not indirect SEM paths.
Each model uses the observations meeting its own inclusion criteria.

\section{Results}
\label{sec:results}
This section follows the estimation strategy described in
Section~\ref{sec:methods} and summarized in
Figure~\ref{fig:estimation-strategy}. 
We first present the national associations (\textbf{H1}) and the connectedness pathway (\textbf{H2}). We then examine their regional extensions (\textbf{H1-Reg} and \textbf{H2-Reg}), the sensitivity to country-by-year effects, and the comparison of regional period averages (\textbf{H3}). Finally, we report the direct contrasts between well-being outcomes (\textbf{H4}). Throughout, we report estimates and confidence intervals to describe the direction, magnitude, and precision of the associations.

\subsection{National associations and connectedness pathways (\textbf{H1} and \textbf{H2})}

To assess \textbf{H1}, we examine the coefficients $\beta_k$ in Equation~\eqref{eq:country}, which relate religious expression to each well-being outcome after accounting for country and year effects and national trends. All three coefficients are positive, and their 95\% confidence intervals exclude zero: 0.317 for \emph{happiness}, 0.491 for \emph{life satisfaction}, and 0.951 for \emph{purpose} (Table~\ref{tab:annual}). Thus, within-country deviations towards more positive religious expression are associated with higher expressed well-being in the same year. \emph{Purpose} has the largest point estimate; whether its association is stronger is assessed through the direct contrasts for \textbf{H4} in Section~\ref{sec:results-h4}.

\begin{table}[htbp]
\centering\small
\caption{Annual religious-expression associations, 2013--2022.}
\label{tab:annual}
\setlength{\tabcolsep}{5pt}
\renewcommand{\arraystretch}{1.25}
\begin{tabular}{@{}llcrr@{}}
\toprule
Geography & Outcome & Estimate [95\% CI] & Place-years & Countries \\
\midrule
Country & \emph{Happiness} & \estci{+0.317}{0.115}{0.519} & 602 & 61 \\
Country & \emph{Life satisfaction} & \estci{+0.491}{0.225}{0.757} & 602 & 61 \\
Country & \emph{Purpose in life} & \estci{+0.951}{0.185}{1.717} & 602 & 61 \\
Region & \emph{Happiness} & \estci{+0.158}{0.059}{0.257} & 8,350 & 60 \\
Region & \emph{Life satisfaction} & \estci{+0.309}{0.182}{0.436} & 8,350 & 60 \\
Region & \emph{Purpose in life} & \estci{+0.708}{0.454}{0.962} & 8,350 & 60 \\
\bottomrule
\end{tabular}
\tabnote{Place and year fixed effects, country-specific linear trends, and country-clustered uncertainty. Predictors and outcomes are standardized within their estimation samples; observations have equal weight.}
\end{table}

For \textbf{H2}, the quantity of interest is the indirect product $a b_k$, combining the religion-to-connectedness coefficient in Equation~\eqref{eq:sems} with the connectedness-to-outcome coefficient in Equation~\eqref{eq:semy}. The estimate $a=0.237$ [0.078, 0.395] indicates that more positive religious expression is associated with higher connectedness one year later, conditional on baseline connectedness and well-being and the specified effects and trends. The $b_k$ coefficients are also positive, with intervals excluding zero for all three outcomes: connectedness is associated with well-being in the following year, conditional on religious expression and the baseline measures (Table~\ref{tab:country-sem}).

The indirect products are 0.060 [0.011, 0.109] for \emph{happiness}, 0.077 [0.004, 0.150] for \emph{life satisfaction}, and 0.159 [$-0.048$, 0.366] for \emph{purpose}. The first two intervals support a positive statistical pathway through connectedness. The \emph{purpose} estimate is larger but less precise, and its interval includes zero. The intervals for the conditional religion--outcome coefficients $c'_k$ and the total associations $c_k=c'_k+a b_k$ also include zero for all three outcomes. The positive indirect components therefore coexist with uncertain total lagged associations; these totals are distinct from the contemporaneous associations used to assess \textbf{H1}.

\begin{table}[htbp]
\centering\small
\caption{Country longitudinal SEM, 2013--2022.}
\label{tab:country-sem}
\setlength{\tabcolsep}{5pt}
\renewcommand{\arraystretch}{1.25}
\begin{tabular}{@{}lccc@{}}
\toprule
Path & \emph{Happiness} & \emph{Life satisfaction} & \emph{Purpose in life} \\
\midrule
$a$ & \estci{+0.237}{0.078}{0.395} & \estci{+0.237}{0.078}{0.395} & \estci{+0.237}{0.078}{0.395} \\
$b_k$ & \estci{+0.255}{0.146}{0.363} & \estci{+0.326}{0.142}{0.510} & \estci{+0.673}{0.038}{1.308} \\
$c'_k$ & \estci{-0.000}{-0.129}{0.129} & \estci{-0.005}{-0.141}{0.131} & \estci{-0.139}{-0.311}{0.032} \\
$a b_k$ & \estci{+0.060}{0.011}{0.109} & \estci{+0.077}{0.004}{0.150} & \estci{+0.159}{-0.048}{0.366} \\
$c_k$ & \estci{+0.060}{-0.077}{0.197} & \estci{+0.072}{-0.073}{0.218} & \estci{+0.020}{-0.193}{0.233} \\
\bottomrule
\end{tabular}
\tabnote{468 place-years in 61 units and 61 countries. Religion at $t$, connectedness at $t+1$, and outcomes at $t+2$; baseline connectedness and all three outcomes controlled. Place/year effects and country trends removed; equal country weight and country-clustered uncertainty. Cells show estimates and 95\% intervals. $a$ is common to all outcomes; $c_k=c'_k+a b_k$. Endogenous intercepts are fixed at zero after residualization. Intervals use country-clustered robust standard errors and the delta method for derived paths.}
\end{table}

The national longitudinal SEM and both regional observed-variable SEMs converged to admissible solutions with positive-definite country-clustered parameter covariance matrices.

The national results support \textbf{H1} for all three outcomes and partially support \textbf{H2}: the connectedness pathway is supported for \emph{happiness} and \emph{life satisfaction}, while the indirect estimate for \emph{purpose} is imprecise.

\subsection{Regional associations and connectedness pathways (\textbf{H1-Reg} and \textbf{H2-Reg})}

To assess \textbf{H1-Reg}, we examine $\beta_k$ in Equation~\eqref{eq:region}, which describes the association within regions after adjustment for region and year effects and national trends. The coefficients are 0.158 for \emph{happiness}, 0.309 for \emph{life satisfaction}, and 0.708 for \emph{purpose}, with all intervals above zero (Table~\ref{tab:annual}). More positive religious expression is therefore associated with higher expressed well-being in the same year within regions, as well as within countries.

For \textbf{H2-Reg}, we examine the indirect products $a b_k$ from Equations~\eqref{eq:sems} and \eqref{eq:semy}, now estimated for regions. The religion-to-connectedness coefficient is 0.052 [$-0.121$, 0.224], and the connectedness-to-outcome coefficients also have intervals including zero. The indirect products are 0.008 [$-0.020$, 0.036] for \emph{happiness}, 0.010 [$-0.025$, 0.045] for \emph{life satisfaction}, and 0.017 [$-0.041$, 0.075] for \emph{purpose} (Table~\ref{tab:region-sem}). Although their point estimates are positive, the intervals do not establish a positive lagged pathway for any outcome. The conditional religion--outcome and total-association intervals likewise include zero.

\begin{table}[htbp]
\centering\small
\caption{Regional longitudinal SEM, 2013--2022.}
\label{tab:region-sem}
\setlength{\tabcolsep}{5pt}
\renewcommand{\arraystretch}{1.25}
\begin{tabular}{@{}lccc@{}}
\toprule
Path & \emph{Happiness} & \emph{Life satisfaction} & \emph{Purpose in life} \\
\midrule
$a$ & \estci{+0.052}{-0.121}{0.224} & \estci{+0.052}{-0.121}{0.224} & \estci{+0.052}{-0.121}{0.224} \\
$b_k$ & \estci{+0.158}{-0.027}{0.344} & \estci{+0.190}{-0.048}{0.427} & \estci{+0.326}{-0.058}{0.710} \\
$c'_k$ & \estci{+0.040}{-0.131}{0.212} & \estci{+0.094}{-0.156}{0.344} & \estci{+0.108}{-0.217}{0.432} \\
$a b_k$ & \estci{+0.008}{-0.020}{0.036} & \estci{+0.010}{-0.025}{0.045} & \estci{+0.017}{-0.041}{0.075} \\
$c_k$ & \estci{+0.049}{-0.145}{0.243} & \estci{+0.104}{-0.177}{0.385} & \estci{+0.124}{-0.251}{0.500} \\
\bottomrule
\end{tabular}
\tabnote{5,122 place-years in 847 units and 60 countries. Religion at $t$, connectedness at $t+1$, and outcomes at $t+2$; baseline connectedness and all three outcomes controlled. Place/year effects and country trends removed; equal country weight and country-clustered uncertainty. Cells show estimates and 95\% intervals. $a$ is common to all outcomes; $c_k=c'_k+a b_k$. Endogenous intercepts are fixed at zero after residualization. Intervals use country-clustered robust standard errors and the delta method for derived paths.}
\end{table}

The regional results support \textbf{H1-Reg} for all three outcomes, but the indirect estimates are too imprecise to support \textbf{H2-Reg}. Positive contemporaneous associations therefore coexist with uncertain lagged relationships within regions. An additional latent regional SEM did not yield interpretable path estimates and provides no further evidence for this hypothesis (Supplementary Section~\ref{app:latent}).

\subsection{Sensitivity of regional associations to country-by-year effects (\textbf{H1-Reg})}

We examine the sensitivity of \textbf{H1-Reg} by comparing $\beta^{\mathrm{CY}}_k$ in Equation~\eqref{eq:region-country-year} with $\beta_k$ in Equation~\eqref{eq:region}. Both models use the same 8,350 region-years from 1,004 regions in 60 countries, with identical weights and standardization. After adjustment for country-by-year effects, the coefficients decrease from 0.158 to 0.129 for \emph{happiness}, from 0.309 to 0.275 for \emph{life satisfaction}, and from 0.708 to 0.695 for \emph{purpose}. All three confidence intervals remain above zero (Table~\ref{tab:country-year-robustness}).

\begin{table}[htbp]
\centering\small
\caption{Regional annual associations under alternative controls for national changes, 2013--2022.}
\label{tab:country-year-robustness}
\setlength{\tabcolsep}{5pt}
\renewcommand{\arraystretch}{1.25}
\begin{tabular}{@{}lcc@{}}
\toprule
Outcome & Country linear trends & Country-by-year effects \\
\midrule
\emph{Happiness} & \estci{+0.158}{0.059}{0.257} & \estci{+0.129}{0.053}{0.205} \\
\emph{Life satisfaction} & \estci{+0.309}{0.182}{0.436} & \estci{+0.275}{0.178}{0.372} \\
\emph{Purpose in life} & \estci{+0.708}{0.454}{0.962} & \estci{+0.695}{0.466}{0.924} \\
\bottomrule
\end{tabular}
\tabnote{Cells show standardized coefficients and 95\% country-clustered
confidence intervals. The first column reproduces the regional estimates
from Table~\ref{tab:annual}. Both specifications use the same 8,350
region-year observations from 1,004 regions in 60 countries, with identical
standardization and equal observation weights. Both include region fixed
effects. The first specification includes common year effects and
country-specific linear trends; the second replaces these with
country-by-year fixed effects.}
\end{table}

The positive associations thus remain after accounting for annual changes shared across regions within each country, supporting \textbf{H1-Reg} under this more flexible specification. Excluding countries with no identifying variation leaves the estimates unchanged and has negligible effects on their confidence intervals (Supplementary Section~\ref{app:regional-identifying-variation}).


\subsection{Associations between regional averages (\textbf{H3})}

To assess \textbf{H3}, we examine $\beta_k$ in Equation~\eqref{eq:static}. This coefficient compares regional averages within countries, rather than annual deviations within regions. Across 733 regions in 43 countries, the coefficients are 0.357 for \emph{happiness}, 0.603 for \emph{life satisfaction}, and 0.889 for \emph{purpose}, with all confidence intervals above zero (Table~\ref{tab:static}). Regions with more positive average religious expression over 2013--2022 also express higher average well-being than other regions in the same country.

\begin{table}[htbp]
\centering\small
\caption{Religious expression and regional mean well-being within countries.}
\label{tab:static}
\setlength{\tabcolsep}{5pt}
\renewcommand{\arraystretch}{1.25}
\begin{tabular}{@{}llcrr@{}}
\toprule
Period & Outcome & Estimate [95\% CI] & Regions & Countries \\
\midrule
2013--2022 & \emph{Happiness} & \estci{+0.357}{0.098}{0.617} & 733 & 43 \\
2013--2022 & \emph{Life satisfaction} & \estci{+0.603}{0.345}{0.862} & 733 & 43 \\
2013--2022 & \emph{Purpose in life} & \estci{+0.889}{0.581}{1.197} & 733 & 43 \\
\bottomrule
\end{tabular}
\tabnote{Country fixed effects, equal total weight per country, and country-clustered uncertainty. At least eight complete annual observations per region during 2013--2022; observed years receive equal weight in each regional mean.}
\end{table}

The complementary static SEM in Equations~\eqref{eq:static-sems} and \eqref{eq:static-semy} examines how connectedness contributes to these associations. Average religious expression is positively associated with average connectedness, $a=0.466$ [0.230, 0.701]. Connectedness is positively associated with each outcome conditional on religion, yielding indirect products $a b_k$ of 0.438 for \emph{happiness}, 0.380 for \emph{life satisfaction}, and 0.284 for \emph{purpose}; all three intervals exclude zero (Table~\ref{tab:static-sem}). These products describe associations among averages for the same period.

\begin{table}[htbp]
\centering\small
\caption{Static regional SEM, 2013--2022.}
\label{tab:static-sem}
\setlength{\tabcolsep}{5pt}
\renewcommand{\arraystretch}{1.25}
\begin{tabular}{@{}lccc@{}}
\toprule
Path & \emph{Happiness} & \emph{Life satisfaction} & \emph{Purpose in life} \\
\midrule
$a$ & \estci{+0.466}{0.230}{0.701} & \estci{+0.466}{0.230}{0.701} & \estci{+0.466}{0.230}{0.701} \\
$b_k$ & \estci{+0.941}{0.858}{1.024} & \estci{+0.816}{0.711}{0.921} & \estci{+0.609}{0.339}{0.879} \\
$c'_k$ & \estci{-0.081}{-0.145}{-0.017} & \estci{+0.223}{0.138}{0.309} & \estci{+0.606}{0.468}{0.743} \\
$a b_k$ & \estci{+0.438}{0.214}{0.662} & \estci{+0.380}{0.180}{0.580} & \estci{+0.284}{0.052}{0.515} \\
$c_k$ & \estci{+0.357}{0.105}{0.609} & \estci{+0.603}{0.353}{0.854} & \estci{+0.889}{0.590}{1.188} \\
\bottomrule
\end{tabular}
\tabnote{733 regions in 43 countries. Country effects removed; equal country weight and country-clustered uncertainty. All measures summarize the same period, without established time ordering. Cells show estimates and 95\% intervals. $a$ is common to all outcomes; $c_k=c'_k+a b_k$. Endogenous intercepts are fixed at zero after residualization. Intervals use country-clustered robust standard errors and the delta method for derived paths.}
\end{table}

The conditional religion--outcome coefficients $c'_k$ in Equation~\eqref{eq:static-semy} show different patterns across outcomes. For \emph{purpose}, the coefficient remains positive at 0.606 [0.468, 0.743], accounting for most of the total association of 0.889. For \emph{happiness}, the conditional coefficient is negative, $-0.081$ [$-0.145$, $-0.017$], alongside a positive indirect product of 0.438 and a positive total association of 0.357. Thus, the positive overall relationship with \emph{happiness} is concentrated in its association through connectedness, while much of the \emph{purpose} association remains after connectedness is included.

The associations between regional averages support \textbf{H3} for all three outcomes. The static SEM describes how these associations are decomposed; it does not test the temporal pathway proposed in \textbf{H2} or \textbf{H2-Reg}.

\subsection{Comparing associations across well-being outcomes (\textbf{H4})}
\label{sec:results-h4}

To assess \textbf{H4}, we compare the \emph{purpose} coefficient $\beta_k$ with the \emph{happiness} and \emph{life satisfaction} coefficients in each association model: Equation~\eqref{eq:country} for countries, Equation~\eqref{eq:region} for regions, and Equation~\eqref{eq:static} for regional averages. A positive difference indicates a stronger association with \emph{purpose}. These direct contrasts, estimated on common samples, assess differences between slopes rather than differences in their separate statistical significance.

At country level, the \emph{purpose}--\emph{happiness} difference is 0.634 [0.033, 1.235], while the \emph{purpose}--\emph{life satisfaction} difference is 0.460 [$-0.071$, 0.991] (Table~\ref{tab:contrasts}). The national results therefore distinguish the \emph{purpose} association from \emph{happiness} but not clearly from \emph{life satisfaction}. At regional level, both differences are positive with intervals excluding zero in the annual models and in the comparison of period averages. The stronger regional \emph{purpose} association is therefore supported by direct comparisons in both designs.

\begin{table}[htbp]
\centering\small
\caption{Direct comparisons of the standardized \emph{purpose} slope with the other outcomes.}
\label{tab:contrasts}
\setlength{\tabcolsep}{5pt}
\renewcommand{\arraystretch}{1.25}
\begin{tabular}{@{}lcc@{}}
\toprule
Design & \emph{Purpose} minus \emph{happiness} & \emph{Purpose} minus \emph{life satisfaction} \\
\midrule
Country annual & \estci{+0.634}{0.033}{1.235} & \estci{+0.460}{-0.071}{0.991} \\
Region annual & \estci{+0.550}{0.352}{0.749} & \estci{+0.399}{0.234}{0.564} \\
Regional means & \estci{+0.532}{0.286}{0.778} & \estci{+0.286}{0.082}{0.490} \\
\bottomrule
\end{tabular}
\tabnote{All models cover 2013--2022. Positive differences indicate a larger \emph{purpose} slope. Common samples and the same controls as the corresponding association models; 95\% country-clustered intervals. The six comparisons are exploratory and not adjusted for multiple testing.}
\end{table}

Repeating the regional contrasts using $\beta^{\mathrm{CY}}_k$ from Equation~\eqref{eq:region-country-year} gives the same ordering. The \emph{purpose} slope exceeds the \emph{happiness} slope by 0.566 [0.392, 0.739] and the \emph{life satisfaction} slope by 0.420 [0.263, 0.577]. The stronger \emph{purpose} association thus remains after accounting for country-by-year effects. These comparisons concern association slopes, not indirect SEM paths.

Overall, \textbf{H4} is partially supported: \emph{purpose} has a stronger association than both other outcomes in the regional designs, while the national evidence supports only the comparison with \emph{happiness}. The contrasts remain exploratory and are not adjusted for multiple testing.
\bigskip 

Taken together, the results show positive contemporaneous associations at both geographic scales and positive associations between regional averages. Evidence for the lagged connectedness pathway is more selective, with support for national \emph{happiness} and \emph{life satisfaction}. Table~\ref{tab:hypothesis-summary} summarizes these findings by hypothesis; Supplementary Section~\ref{app:figures} presents the coefficient figures and estimated path diagrams.

\begin{table}[htbp]
\centering\small
\caption{Summary of evidence for the hypotheses.}
\label{tab:hypothesis-summary}
\setlength{\tabcolsep}{4pt}
\renewcommand{\arraystretch}{1.2}
\begin{tabularx}{\linewidth}{@{}l p{.23\linewidth} l X@{}}
\toprule
Hypothesis & Question & Support & Evidence \\
\midrule
\textbf{H1} & National contemporaneous associations & Yes & Positive associations for all three outcomes; all intervals exclude zero. \\
\textbf{H2} & National connectedness pathway & Partial & Positive indirect associations for \emph{happiness} and \emph{life satisfaction}; the indirect estimate for \emph{purpose} is imprecise. \\
\textbf{H1-Reg} & Regional contemporaneous associations & Yes & Positive associations for all three outcomes, also with country-by-year effects. \\
\textbf{H2-Reg} & Regional connectedness pathway & No & Imprecise indirect estimates: all confidence intervals include zero. \\
\textbf{H3} & Associations between regional averages & Yes & Positive associations for all three outcomes within countries. \\
\textbf{H4} & Stronger association with \emph{purpose} & Partial & Both contrasts are positive in the regional designs. Nationally, \emph{purpose} exceeds \emph{happiness}; the estimated contrast with \emph{life satisfaction} is imprecise. \\
\bottomrule
\end{tabularx}
\tabnote{Support summarizes the direction of the estimates and their reported 95\% confidence intervals within the specified models. ``No'' means that the predicted association is not established by these estimates, not that its absence is demonstrated. Partial support indicates support for some outcomes or comparisons. Assessments of \textbf{H2} and \textbf{H2-Reg} concern indirect associations, not total associations or differences between indirect paths. The six \textbf{H4} contrasts are exploratory and unadjusted for multiple testing.}
\end{table}
\FloatBarrier


\section{Limitations}
\label{sec:limitations}

The interpretation of these results follows from the type of observation that HFGI provides. The scores used here summarize public expressions relative to all observed posts in each geographic unit and year, combining the prevalence and valence of expression. The models do not separate associations attributable to these two components. They do not directly measure the prevalence of religious belief or the average well-being of residents \citep{iacus2026hfgi}. Survey responses and online posts also arise through different processes: surveys elicit answers to specified questions, while posting depends on what people choose to communicate and to whom. Surveys can be affected by nonresponse, recall, question wording, and social desirability; online expression is shaped by audiences, self-presentation, and the salience of events. Agreement between the sources can be informative without implying that they measure identical quantities, and disagreement does not necessarily establish by itself that either source is invalid \citep{oecd2025wellbeing}.

Selection occurs at several stages, from access to and use of the platform to public posting and the availability of geographic information \citep{olteanu2019social}. Geolocated users can differ even from other users of the same platform \citep{sloan2015geotagging}. Consequently, the observed stream need not reproduce the demographic or geographic composition of the population. 

Selection continues within the observed stream because people differ in whether they discuss religion, social relationships, or their well-being, and in whether they express those experiences positively or negatively. A change in an indicator can therefore reflect a change in experience, a change in willingness to express it, or a change in the composition of those posting. Religious norms, perceived stigma, and public events may affect these processes differently across places. An increase in a religious-expression score can reflect changes in how often religious dimensions are expressed, in the valence of those expressions, or both; it cannot be read as an increase in the proportion of religious residents.  

Geographic and temporal aggregation introduce further constraints. Posting volumes, geolocation practices, and platform access can change over time, and geographic attribution does not necessarily identify a poster's usual residence. Minimum tweet counts and coverage thresholds reduce dependence on very sparse cells, but do not correct the composition of the observed population. Region effects absorb persistent additive differences, while country-by-year effects absorb annual shifts shared by regions within each country. These adjustments leave open differential changes in regional user composition, expression practices, and exposure to national events. The persistence of the regional association under these controls therefore supports its robustness to those particular adjustments, rather than eliminating selection or measurement bias.

Automated classification adds uncertainty of its own. The published HFGI validation and cross-linguistic model assessment support the measurement approach \citep{iacus2026hfgi,nasuto2026language}, while ambiguity, irony, and culturally specific expression can still affect particular indicators. The common posting source and classification framework may generate correlated errors across religious expression, connectedness, and well-being.

\section{Discussion and conclusions}
\label{sec:discussion}

The study provides evidence for a national statistical pathway linking religious expression to subsequent \emph{happiness} and \emph{life satisfaction} through social connectedness. This is the main contribution of the longitudinal analysis: religious expression is associated with connectedness one year later, which is in turn associated with these outcomes in the following year, conditional on baseline measures, fixed effects, and trends. The broader results show positive contemporaneous associations across all three well-being dimensions and positive associations between regional averages. Together, these findings distinguish a connectedness pathway from the wider pattern of religious-expression associations summarized in Table~\ref{tab:hypothesis-summary}.

The national pathway gives a territorial perspective on the social relationships emphasized in earlier research. \citet{lim2010religion} link religious-service attendance and congregational friendships to \emph{life satisfaction}, with the friendship association depending on a strong religious identity. \citet{diener2011religion} identify social support, respect, and meaning as statistical intermediaries in the relationship between religiosity and well-being. Our analysis examines a broader form of connection: belonging to society, lower loneliness, trust in others, and feeling trusted. The positive national indirect associations indicate that this broader connectedness measure is relevant to the relationship between religious expression and later well-being. They extend the examination of social connection to repeated territorial expressions, while the individual and congregational processes studied in earlier work remain distinct from our aggregate measures.

\emph{Purpose} adds a complementary dimension to this account. Its association with religious expression exceeds those of \emph{happiness} and \emph{life satisfaction} in both regional designs, and this ordering remains under country-by-year controls. In the static regional SEM, a substantial \emph{purpose} association also remains after connectedness is included. These findings make the distinction between relational and existential resources useful for interpretation. \citet{Becchetti2025} examines social interaction and meaning in life among several channels connecting religiosity to \emph{life satisfaction}. \citet{steger2005meaning} find that meaning in life statistically mediates relationships between religiousness and well-being. Here, \emph{purpose} is an outcome, allowing us to examine its association with religious expression alongside affective and evaluative well-being.
The findings suggest that the relationship between religious expression and well-being may involve distinct social and existential processes. At the national level, social connectedness provides a statistical pathway linking religious expression to subsequent \emph{happiness} and \emph{life satisfaction}. The stronger regional association with \emph{purpose}, which remains substantial after accounting for connectedness, is compatible with a more personal dimension of religious experience, through which beliefs and comfort may provide meaning and direction \citep{prinzig2023}. This interpretation remains tentative: the imprecise longitudinal indirect estimate for \emph{purpose} does not establish that its association with religious expression is independent of social relationships.

Emotional processes provide another substantive direction for investigation. \citet{VanCappellenEtAl2016} report statistical mediation through self-transcendent positive emotions, including awe and gratitude, in two cross-sectional studies of religion or spirituality and well-being. 
In the present work, unmeasured emotional processes and shared measurement variation remain possible contributors to the conditional associations.
In future extensions, emotional experiences may be examined as mediators alongside social connection.

The geographic comparisons clarify which relationships the data support at each scale. Positive contemporaneous associations appear within both countries and regions, and the regional associations persist after adjustment for country-by-year effects. The regional lagged estimates remain imprecise, whereas the national indirect intervals support the pathway for \emph{happiness} and \emph{life satisfaction}. This difference in precision does not establish a difference between the pathways, but it identifies the national model as the clearest evidence for the proposed sequence in these data. Contextual research offers a reason to investigate geographic variation: \citet{stavrova2013religious} find stronger individual religiosity--well-being associations where religiosity is socially normative, partly associated with greater respect for religious individuals. Our comparisons add evidence on temporal and geographic patterns of expression, even if they do not directly test that social-norm explanation.

The findings also extend research on religion in online expression. \citet{RitterEtAl2014} find more positive and less negative emotional language among Christian than atheist Twitter users, with social-relationship language accounting for part of the difference in expressed \emph{happiness}. Our contribution is to examine religious-expression scores in repeated geographic observations and distinguish \emph{happiness}, \emph{life satisfaction}, and \emph{purpose}. The combination of temporal models, regional comparisons, and direct outcome contrasts shows where connectedness contributes to these associations and where the evidence remains uncertain.

The magnitudes should be interpreted in relation to these measures and units. \citet{Prati2024} reports very small or negligible associations between religious indicators and well-being in large individual-level survey analyses, questioning their practical importance. Our standardized territorial expression coefficients describe a different relationship and cannot be interpreted as individual well-being gains or compared directly with those survey effect sizes. The direct contrasts establish a stronger regional association with \emph{purpose}, while the imprecise total national lagged associations leave the practical magnitude of the proposed sequence unresolved.

Across countries and regions, online expression reveals patterns consistent with established research on religion and well-being. By combining broad geographic coverage with annual observations over a decade, this study distinguishes temporally ordered associations from contemporaneous relationships and persistent differences between regions. The clearest longitudinal evidence concerns the national connectedness pathways for \emph{happiness} and \emph{life satisfaction}, while the regional comparisons highlight the stronger association with \emph{purpose}. These findings concern public expressions among the observed posting population and do not establish causal effects on individual well-being.

\section*{Abbreviations}

\begin{description}
\item[CFI] Comparative fit index.
\item[CI] Confidence interval.
\item[CY] Country-by-year.
\item[HFGI] Human Flourishing Geographic Index.
\item[NAISS] National Academic Infrastructure for Supercomputing in Sweden.
\item[RMSEA] Root mean square error of approximation.
\item[SEM] Structural equation model.
\end{description}

\section*{Declarations}

\subsection*{Availability of data and materials}

The model input data, analysis code, and supporting documentation are available at \url{https://github.com/siacus/religion-hfgi}. The repository provides a single analysis dataset and documents indicator construction, model-specific scaling, sample selection, estimation, and sensitivity analyses. It reproduces the reported numerical results, coefficient figures, and supplementary latent-model diagnostics, and supplies the diagram sources.

\subsection*{Competing interests}

The authors declare that they have no competing interests.

\subsection*{Funding}

This research received no specific grant from any funding agency in the public, commercial, or not-for-profit sectors.

\subsection*{Authors' contributions}

Stefano M. Iacus: Conceptualization, methodology, software, data curation, formal analysis, visualization, and writing---original draft, review and editing. Giuseppe Porro: Conceptualization and writing---original draft, review and editing. Haodong Qi: Writing---original draft, review and editing.

\subsection*{Acknowledgements}

The authors acknowledge the computational resources provided by the National Academic Infrastructure for Supercomputing in Sweden (NAISS).

\subsection*{Use of AI-assisted tools}

OpenAI Codex was used to assist with text correction and the development, verification, and debugging of analysis scripts. The authors take full responsibility for the manuscript, code, and results.


\FloatBarrier
\bibliographystyle{plainnat}
\bibliography{references}
\clearpage
\appendix
\section*{Supplementary material}
\addcontentsline{toc}{section}{Supplementary material}
\renewcommand{\thesection}{S\arabic{section}}
\setcounter{figure}{0}
\setcounter{table}{0}
\setcounter{equation}{0}
\renewcommand{\thefigure}{S\arabic{figure}}
\renewcommand{\thetable}{S\arabic{table}}
\renewcommand{\theequation}{S\arabic{equation}}
\renewcommand{\theHsection}{supp.\arabic{section}}
\renewcommand{\theHfigure}{supp.\arabic{figure}}
\renewcommand{\theHtable}{supp.\arabic{table}}
\renewcommand{\theHequation}{supp.\arabic{equation}}

\section{Data preparation and estimation details}
\label{app:data-preparation}

\subsection{Definition of the dimension scores}

Before standardization, each dimension score is the sum of the scores assigned to posts expressing that dimension, divided by the total number of posts in the geographic unit and year. Posts in which the dimension is not detected contribute zero to this sum and remain in the denominator. Equivalently, the score is the share of posts expressing the dimension multiplied by their mean expression score. The national US indicators are constructed by summing the corresponding score totals and post counts across the 50 states and the District of Columbia before calculating each national ratio. The composites and standardization described below are applied after this geographic aggregation.

\subsection{Coverage and inclusion criteria}

All data preparation, index standardization, and analyses are restricted to 2013--2022. Observations from 2023 cover only January--June and are excluded. We use annual aggregate expression scores, together with the number of posts for each geographic unit and the share expressing each dimension. The approximate number of posts expressing a dimension is recovered by multiplying that share by the total number of posts and rounding to the nearest integer. A geographic unit and year must have at least 1,000 posts, with at least 30 posts expressing each required dimension.

Records without geographic identifiers or with missing, non-finite, or out-of-range scores are excluded; the shares expressing each dimension must also lie within their admissible range. Duplicate records for the same geographic unit, year, and dimension are resolved before constructing the panels. For religious-expression and outcome measures, records with a country name are preferred, followed by those with the largest total number of posts. For connectedness measures, the eligible record with the largest total number of posts is retained.

Annual association samples require complete religious-expression and outcome measures and at least four eligible years per geographic unit. They contain 602 country-years in 61 countries and 8,350 region-years in 1,004 regions and 60 countries. The longitudinal pathway samples additionally require all connectedness dimensions and observations at $t$, $t+1$, and $t+2$. Units must contribute at least three eligible baseline years. These requirements yield 468 country-years in 61 countries and 5,122 region-years in 847 regions and 60 countries. Country-years and region-years in these pathway samples refer to eligible baseline observations with their corresponding follow-up measures.

\subsection{Standardization of the annual measures}

For the annual analyses, the four religious-expression dimensions are first standardized separately within the country or regional source panel for 2013--2022, then averaged with equal item weights. All four dimensions must be observed. In the same-year association models, this composite and each outcome are subsequently standardized within the estimation sample.

For the longitudinal pathway analyses, the four connectedness dimensions are standardized using weights that give each country equal total weight. Loneliness is reversed, the four dimensions are averaged with equal item weights, and the resulting index is standardized again. The religious-expression composite and all three outcomes are also standardized using these weights in the complete annual pathway panel, before constructing the lagged sample. The SEM coefficients thus refer to scaled observed measures; the residuals obtained after removing fixed effects and trends are not standardized again. The weighting used in model estimation is described in the main methods section.

\subsection{Construction of regional period averages}

For the static analysis, regional means are calculated from years with complete religious-expression, connectedness, and outcome measures during 2013--2022. Each region must contribute at least eight annual observations, and each year receives equal weight in its period mean. Countries must contain at least two eligible regions. Regional item means are standardized with equal total weight per country before forming and restandardizing the two equal-item indices; loneliness is reversed. The final sample comprises 733 regions in 43 countries. These averages describe differences between regions over a shared period, whereas the annual models describe variation over time.

Previous research provides evidence on the cross-linguistic performance of the language models underlying this approach. \citet{nasuto2026language} document transfer of topic recognition across languages and improvements in more nuanced classification through multilingual fine-tuning. This evidence complements the indicator construction and validation reported in the HFGI data descriptor \citep{iacus2026hfgi}. 
\subsection{Covariance structure and model degrees of freedom}
\label{app:sem-specification}

The national longitudinal, regional longitudinal, and static regional observed-variable SEMs have saturated covariance structures: they impose no restrictions on the modeled covariances that could independently test the proposed paths. Their covariance-structure degrees of freedom are therefore zero. The four fixed intercepts produce four nominal model degrees of freedom in \texttt{lavaan}, but their zero means follow from the preceding residualization. Global fit is consequently not used to validate the substantive path structure.

\subsection{Sensitivity to countries without identifying variation}
\label{app:regional-identifying-variation}

A country-year containing only one observed region provides no variation for the regional comparison after its country-by-year effect is removed. We therefore examine the variation in religious expression remaining after adjustment for region and country-by-year effects in Equation~\eqref{eq:region-country-year}, and repeat the model excluding countries with none. We retain the original standardization, equal observation weights, and country-clustered standard errors. This check assesses whether including countries that contribute no identifying variation affects the estimated uncertainty. It concerns the contemporaneous regional association, rather than the longitudinal connectedness pathway.

The country-by-year specification retains 74.4\% of the baseline residual predictor sum of squares. Of 572 country-year cells, 94 contain a single observed region; 54 countries contribute residual identifying variation. Excluding the other six countries reduces the sample from 8,350 region-years in 1,004 regions and 60 countries to 8,307 region-years in 997 regions and 54 countries. The point estimates are unchanged. The resulting 95\% confidence intervals remain positive: [0.053, 0.205] for \emph{happiness}, [0.178, 0.373] for \emph{life satisfaction}, and [0.465, 0.925] for \emph{purpose}. Table~\ref{tab:country-year-robustness} reports the comparison on the full sample.

\subsection{Religious affiliation and fixed effects}
\label{app:affiliation-adjustment}

The fixed effects also account for stable differences in religious affiliation, which is conceptually distinct from religious expression. National affiliation shares from Pew's 2010/2020 composition estimates, processed by Our World in Data, and Christian-tradition estimates referring to 2010 describe identification with religious groups \citep{hackett2025composition,owid2025affiliation,pew2011christianity}. A selected year's national shares are time-invariant: country effects absorb them in the national and static regional models, and region effects absorb them in the annual regional models. The corresponding effects are removed before fitting the SEMs, so separate main coefficients for these shares are not identified. This adjustment covers stable affiliation differences, rather than changes in affiliation over time.

\section{Coefficient figures and estimated path diagrams}
\label{app:figures}

The following figures reproduce the country, regional, and static
regional results. The path diagrams display point estimates, with full
uncertainty reported in the main tables. The common $a$ coefficient is
repeated for readability. For the annual models, baseline connectedness
and all three baseline outcomes enter every equation; they are omitted
from these diagrams. The three outcome residuals may covary in each SEM.
Arrows describe the estimated statistical relationships.

\begin{figure}[p]
\centering
\includegraphics[width=.92\linewidth]{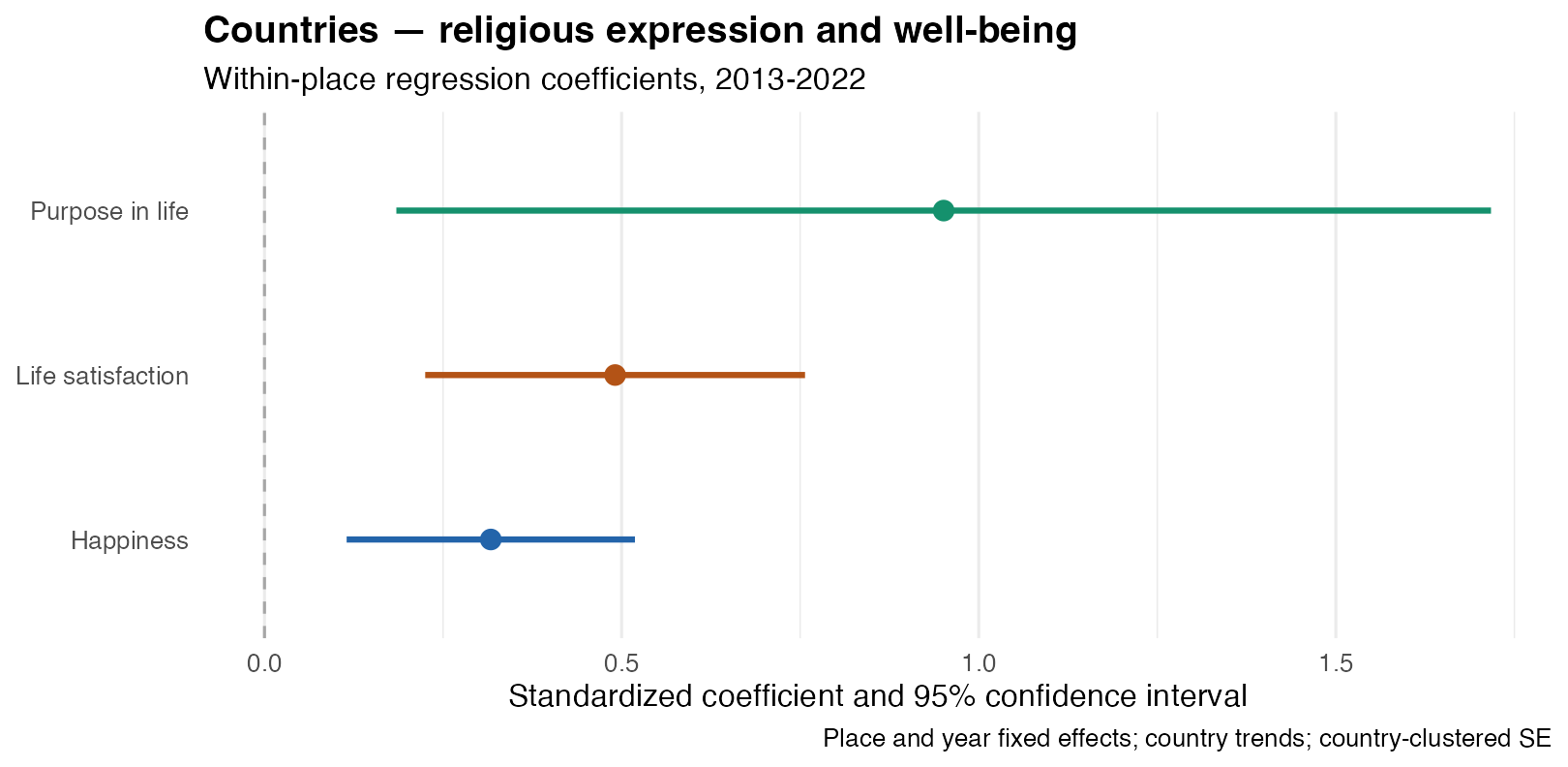}
\par\bigskip
\includegraphics[width=\linewidth]{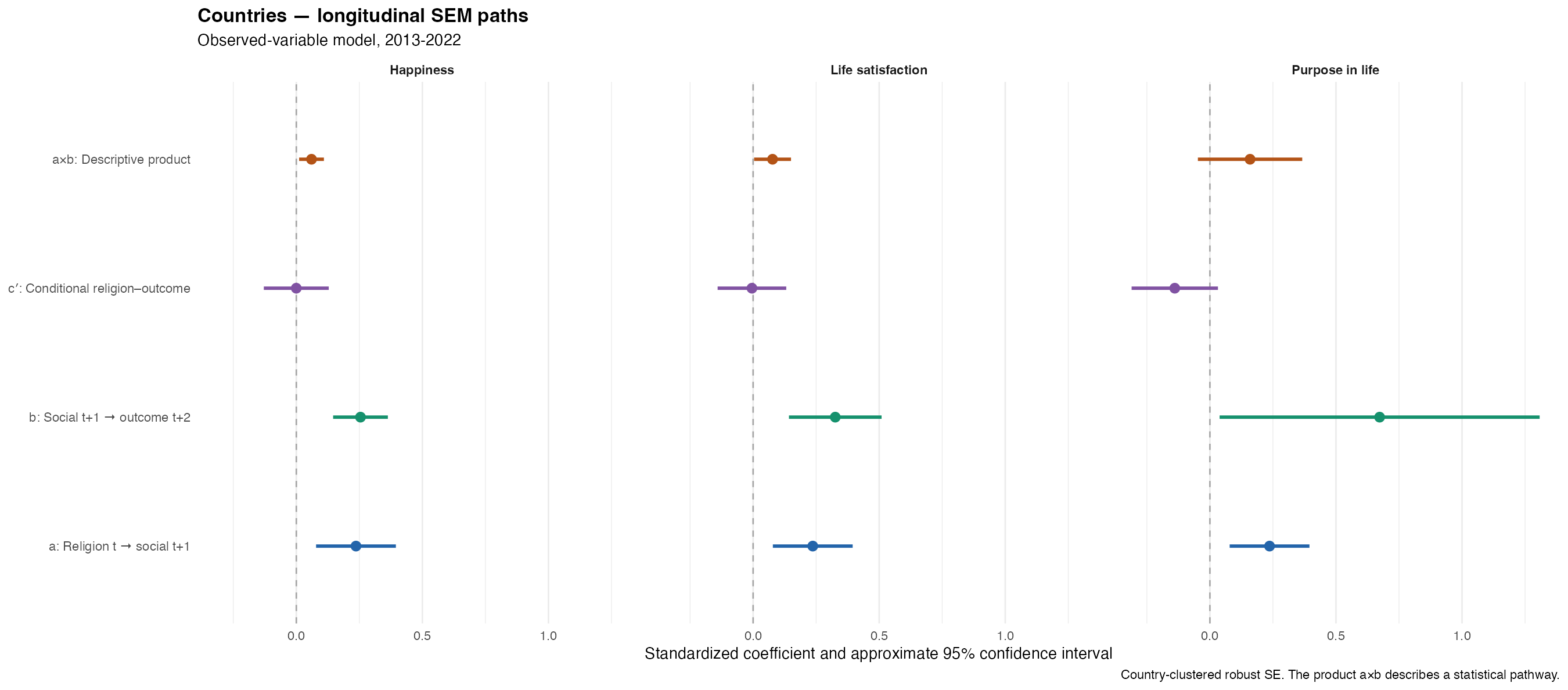}
\caption{Country-level same-year associations and longitudinal SEM paths.
The panels use different analysis samples. SEM intervals are exploratory.}
\label{fig:country-coefficients}
\end{figure}

\begin{figure}[p]
\centering
\begin{tikzpicture}[>=Stealth,box/.style={draw,rounded corners,align=center,font=\small,text width=29mm,minimum height=12mm}]
\node[box] (r) at (0,0) {Religion\\$R_t$};
\node[box] (s) at (4.6,1.7) {Connectedness\\$S_{t+1}$};
\node[box] (y) at (9.2,0) {\emph{Happiness}\\$t+2$};
\draw[->] (r)--node[above left,font=\small]{$a=+0.237$}(s);
\draw[->] (s)--node[above right,font=\small]{$b_k=+0.255$}(y);
\draw[->] (r)--node[below,font=\small]{$c'_k=-0.000$}(y);
\end{tikzpicture}\par\bigskip
\begin{tikzpicture}[>=Stealth,box/.style={draw,rounded corners,align=center,font=\small,text width=29mm,minimum height=12mm}]
\node[box] (r) at (0,0) {Religion\\$R_t$};
\node[box] (s) at (4.6,1.7) {Connectedness\\$S_{t+1}$};
\node[box] (y) at (9.2,0) {\emph{Life satisfaction}\\$t+2$};
\draw[->] (r)--node[above left,font=\small]{$a=+0.237$}(s);
\draw[->] (s)--node[above right,font=\small]{$b_k=+0.326$}(y);
\draw[->] (r)--node[below,font=\small]{$c'_k=-0.005$}(y);
\end{tikzpicture}\par\bigskip
\begin{tikzpicture}[>=Stealth,box/.style={draw,rounded corners,align=center,font=\small,text width=29mm,minimum height=12mm}]
\node[box] (r) at (0,0) {Religion\\$R_t$};
\node[box] (s) at (4.6,1.7) {Connectedness\\$S_{t+1}$};
\node[box] (y) at (9.2,0) {\emph{Purpose in life}\\$t+2$};
\draw[->] (r)--node[above left,font=\small]{$a=+0.237$}(s);
\draw[->] (s)--node[above right,font=\small]{$b_k=+0.673$}(y);
\draw[->] (r)--node[below,font=\small]{$c'_k=-0.139$}(y);
\end{tikzpicture}\par\bigskip
\caption{Country longitudinal SEM paths. Point estimates only; see the corresponding table for uncertainty.}
\label{fig:country-paths}
\end{figure}

\clearpage

\begin{figure}[p]
\centering
\includegraphics[width=.92\linewidth]{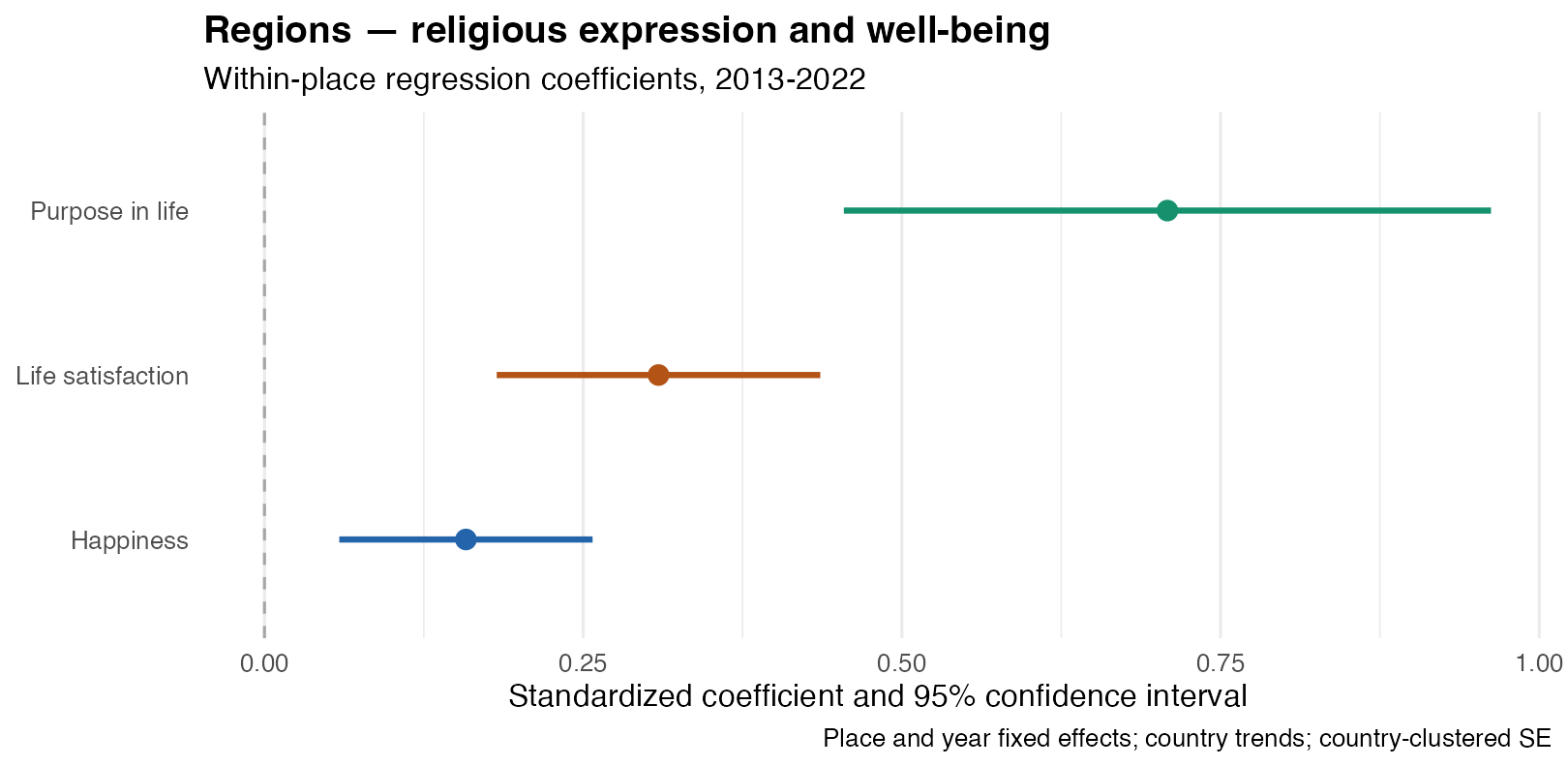}
\par\bigskip
\includegraphics[width=\linewidth]{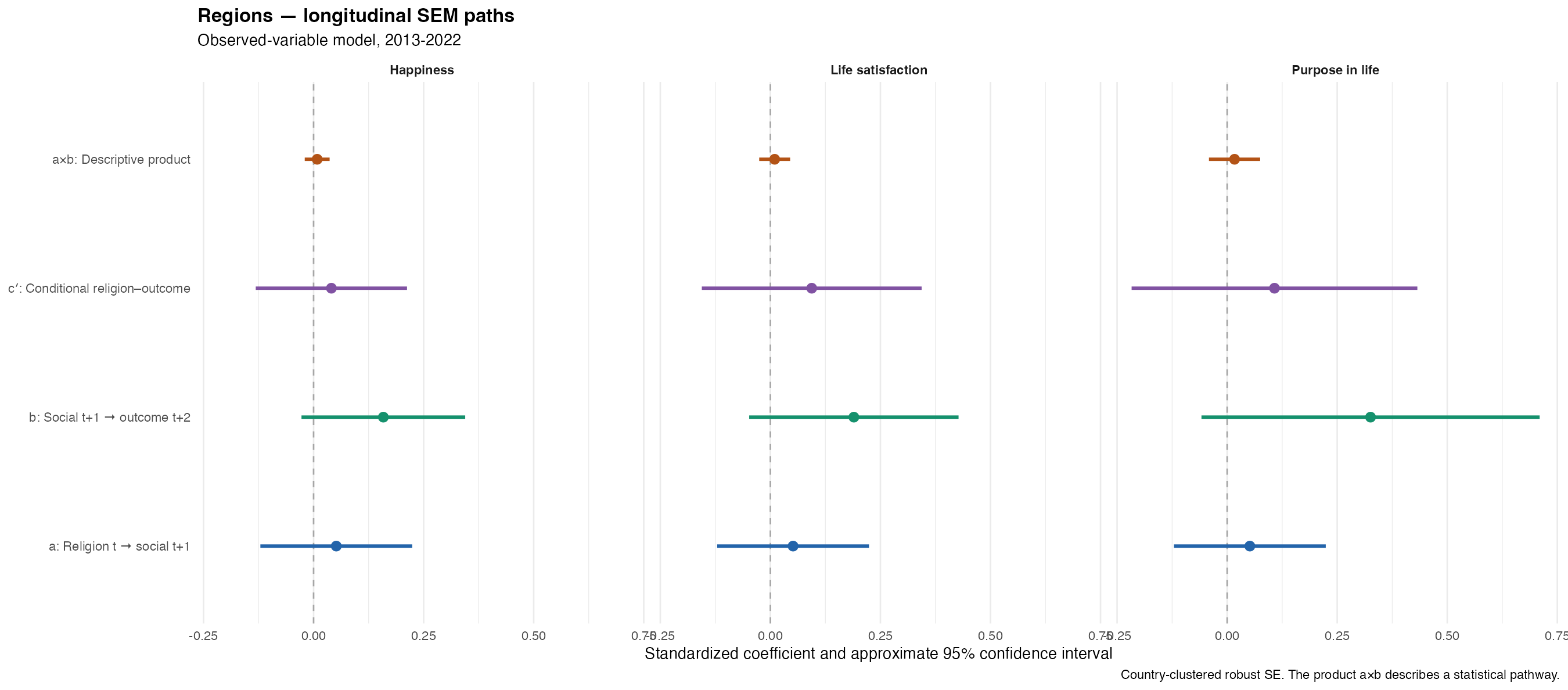}
\caption{Regional same-year associations and longitudinal SEM paths.
All focal regional SEM intervals include zero.}
\label{fig:regional-coefficients}
\end{figure}

\begin{figure}[p]
\centering
\begin{tikzpicture}[>=Stealth,box/.style={draw,rounded corners,align=center,font=\small,text width=29mm,minimum height=12mm}]
\node[box] (r) at (0,0) {Religion\\$R_t$};
\node[box] (s) at (4.6,1.7) {Connectedness\\$S_{t+1}$};
\node[box] (y) at (9.2,0) {\emph{Happiness}\\$t+2$};
\draw[->] (r)--node[above left,font=\small]{$a=+0.052$}(s);
\draw[->] (s)--node[above right,font=\small]{$b_k=+0.158$}(y);
\draw[->] (r)--node[below,font=\small]{$c'_k=+0.040$}(y);
\end{tikzpicture}\par\bigskip
\begin{tikzpicture}[>=Stealth,box/.style={draw,rounded corners,align=center,font=\small,text width=29mm,minimum height=12mm}]
\node[box] (r) at (0,0) {Religion\\$R_t$};
\node[box] (s) at (4.6,1.7) {Connectedness\\$S_{t+1}$};
\node[box] (y) at (9.2,0) {\emph{Life satisfaction}\\$t+2$};
\draw[->] (r)--node[above left,font=\small]{$a=+0.052$}(s);
\draw[->] (s)--node[above right,font=\small]{$b_k=+0.190$}(y);
\draw[->] (r)--node[below,font=\small]{$c'_k=+0.094$}(y);
\end{tikzpicture}\par\bigskip
\begin{tikzpicture}[>=Stealth,box/.style={draw,rounded corners,align=center,font=\small,text width=29mm,minimum height=12mm}]
\node[box] (r) at (0,0) {Religion\\$R_t$};
\node[box] (s) at (4.6,1.7) {Connectedness\\$S_{t+1}$};
\node[box] (y) at (9.2,0) {\emph{Purpose in life}\\$t+2$};
\draw[->] (r)--node[above left,font=\small]{$a=+0.052$}(s);
\draw[->] (s)--node[above right,font=\small]{$b_k=+0.326$}(y);
\draw[->] (r)--node[below,font=\small]{$c'_k=+0.108$}(y);
\end{tikzpicture}\par\bigskip
\caption{Regional longitudinal SEM paths. Point estimates only; see the corresponding table for uncertainty.}
\label{fig:region-paths}
\end{figure}

\clearpage

\begin{figure}[p]
\centering
\includegraphics[width=.92\linewidth]{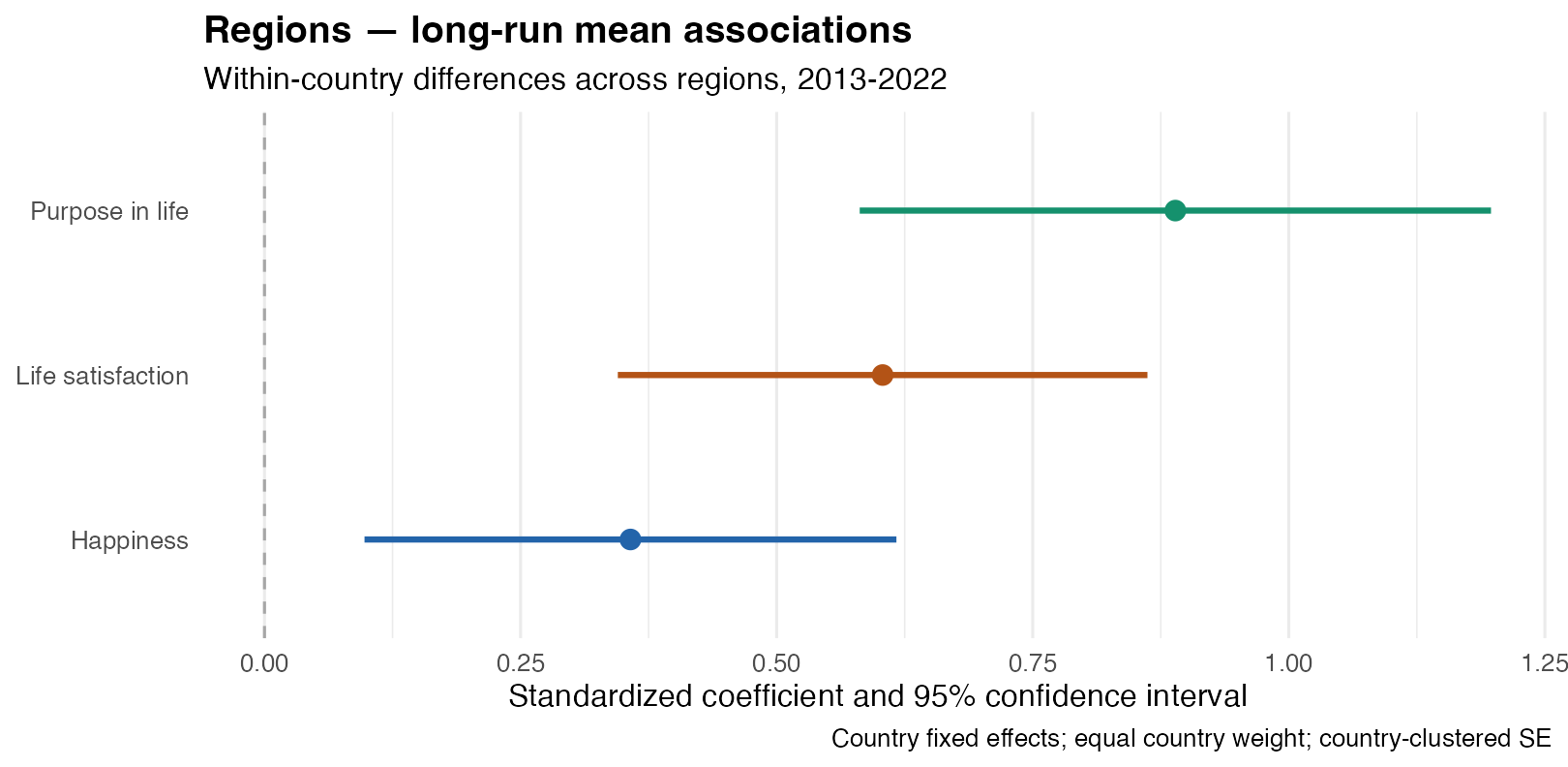}
\par\bigskip
\includegraphics[width=\linewidth]{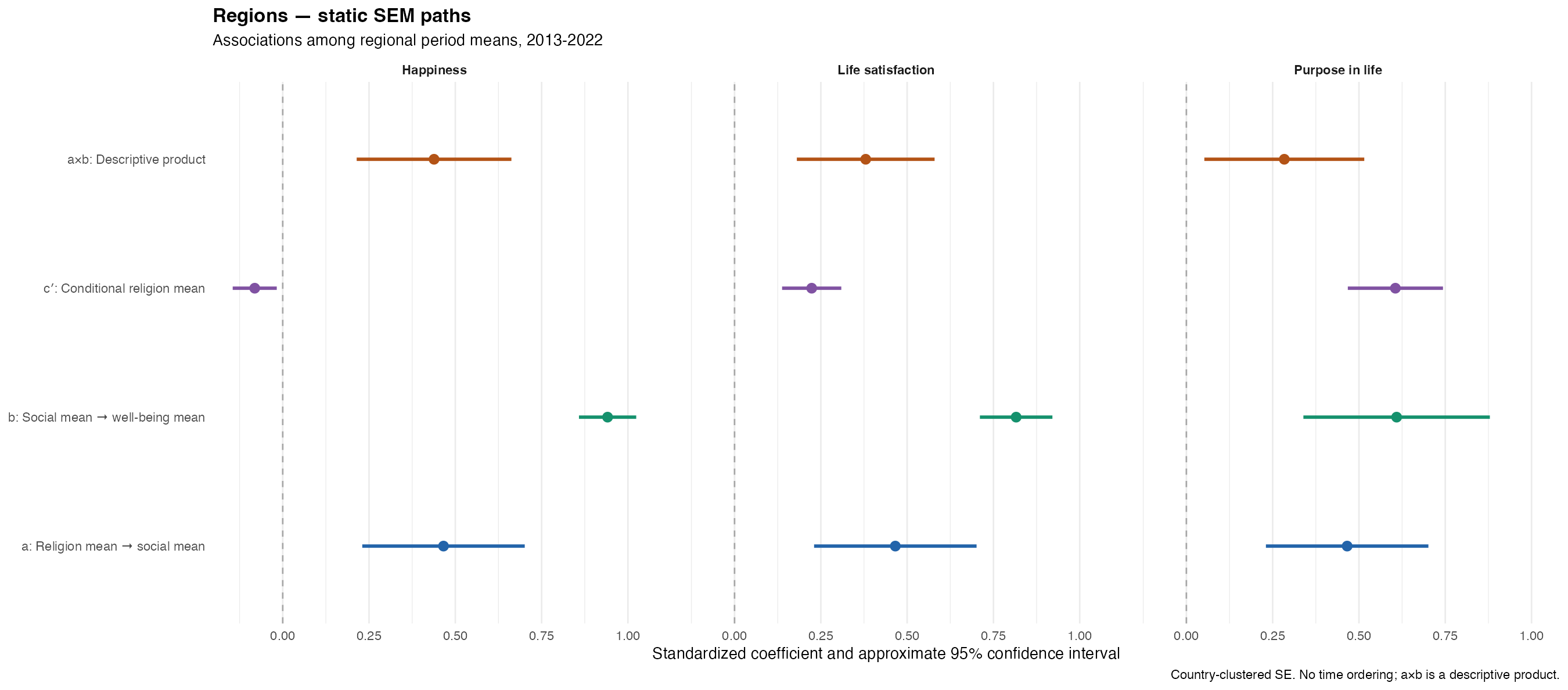}
\caption{Associations and SEM paths for regional period means,
2013--2022. All measures cover the same period.}
\label{fig:static-coefficients}
\end{figure}

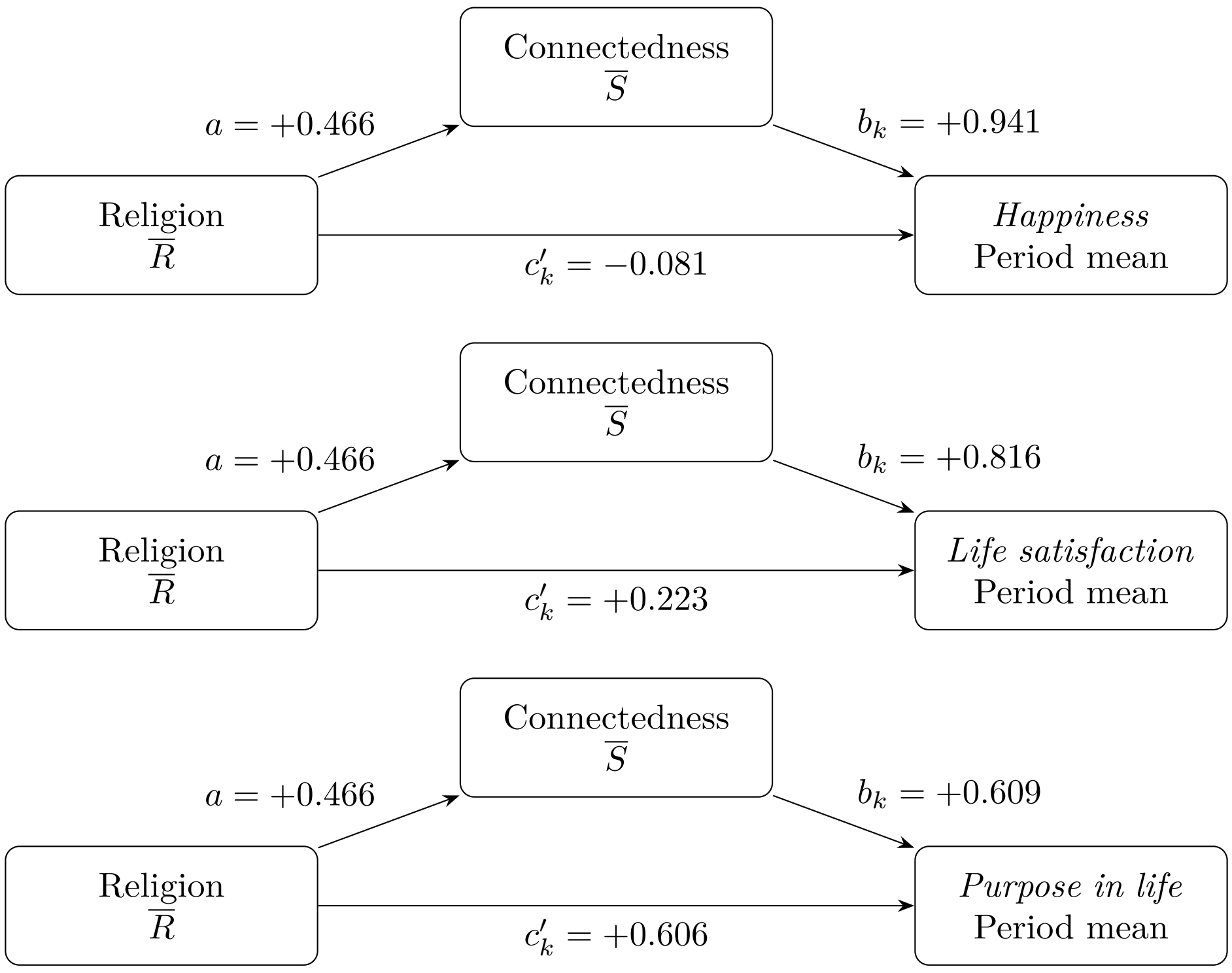
\begin{figure}[p]
\centering
\begin{tikzpicture}[>=Stealth,box/.style={draw,rounded corners,align=center,font=\small,text width=29mm,minimum height=12mm}]
\node[box] (r) at (0,0) {Religion\\$\overline R$};
\node[box] (s) at (4.6,1.7) {Connectedness\\$\overline S$};
\node[box] (y) at (9.2,0) {\emph{Happiness}\\Period mean};
\draw[->] (r)--node[above left,font=\small]{$a=+0.466$}(s);
\draw[->] (s)--node[above right,font=\small]{$b_k=+0.941$}(y);
\draw[->] (r)--node[below,font=\small]{$c'_k=-0.081$}(y);
\end{tikzpicture}\par\bigskip
\begin{tikzpicture}[>=Stealth,box/.style={draw,rounded corners,align=center,font=\small,text width=29mm,minimum height=12mm}]
\node[box] (r) at (0,0) {Religion\\$\overline R$};
\node[box] (s) at (4.6,1.7) {Connectedness\\$\overline S$};
\node[box] (y) at (9.2,0) {\emph{Life satisfaction}\\Period mean};
\draw[->] (r)--node[above left,font=\small]{$a=+0.466$}(s);
\draw[->] (s)--node[above right,font=\small]{$b_k=+0.816$}(y);
\draw[->] (r)--node[below,font=\small]{$c'_k=+0.223$}(y);
\end{tikzpicture}\par\bigskip
\begin{tikzpicture}[>=Stealth,box/.style={draw,rounded corners,align=center,font=\small,text width=29mm,minimum height=12mm}]
\node[box] (r) at (0,0) {Religion\\$\overline R$};
\node[box] (s) at (4.6,1.7) {Connectedness\\$\overline S$};
\node[box] (y) at (9.2,0) {\emph{Purpose in life}\\Period mean};
\draw[->] (r)--node[above left,font=\small]{$a=+0.466$}(s);
\draw[->] (s)--node[above right,font=\small]{$b_k=+0.609$}(y);
\draw[->] (r)--node[below,font=\small]{$c'_k=+0.606$}(y);
\end{tikzpicture}\par\bigskip
\caption{Static regional SEM paths, 2013--2022. Point estimates only; see the corresponding table for uncertainty.}
\label{fig:static-paths}
\end{figure}

\clearpage

\section{Latent multilevel SEM: specification and diagnostics}
\label{app:latent}

We also attempted a two-level latent SEM on the regional lagged panel.
Before fitting, each observed measure was residualized for country and
year fixed effects and country-specific linear trends. For level
$\ell\in\{W,B\}$, denoting annual within-region deviations and the
between-region component of the residualized variables, respectively,
the measurement equations are
\begin{align}
X^R_{j\ell}&=\lambda^R_{j\ell}R_\ell+e^R_{j\ell},\\
X^S_{m\ell}&=\lambda^S_{m\ell}S_\ell+e^S_{m\ell}.
\end{align}
Here $j$ indexes \emph{belief in an afterlife} (\texttt{afterlife}), \emph{belief in God} (\texttt{believegod}), \emph{feeling loved by God} (\texttt{lovedgod}), and \emph{religious comfort} (\texttt{relcomfort}); $m$ indexes \emph{belonging} (\texttt{belonging}), reversed \emph{loneliness} (\texttt{loneliness}), \emph{trust in others} (\texttt{trust}), and \emph{feeling trusted by others} (\texttt{trusted}).
Here $X^R$ and $X^S$ denote the observed religious-expression and connectedness indicators, $R_\ell$ and $S_\ell$ their latent factors, $\lambda^R$ and $\lambda^S$ the factor loadings, and $e^R$ and $e^S$ the measurement residuals. The structural equations are
\begin{align}
S_\ell&=a_\ell R_\ell+u_\ell,\\
Y^{(k)}_\ell&=b_{k\ell}S_\ell+c'_{k\ell}R_\ell+v^{(k)}_\ell.
\end{align}
The terms $u_\ell$ and $v^{(k)}_\ell$ are the connectedness and outcome equation residuals. Religion indicators are measured at $t$, connectedness indicators at
$t+1$, and outcomes at $t+2$. At each level, $a_\ell b_{k\ell}$ would
be the indirect product and $c'_{k\ell}+a_\ell b_{k\ell}$ the total
modeled association. These coefficients are defined to explain the
specification; they are not interpreted. The between-region component
is specific to this lagged residualized sample and is distinct from
the static period means. This diagnostic model omits the baseline
connectedness and well-being controls used in the observed-index
longitudinal SEMs.

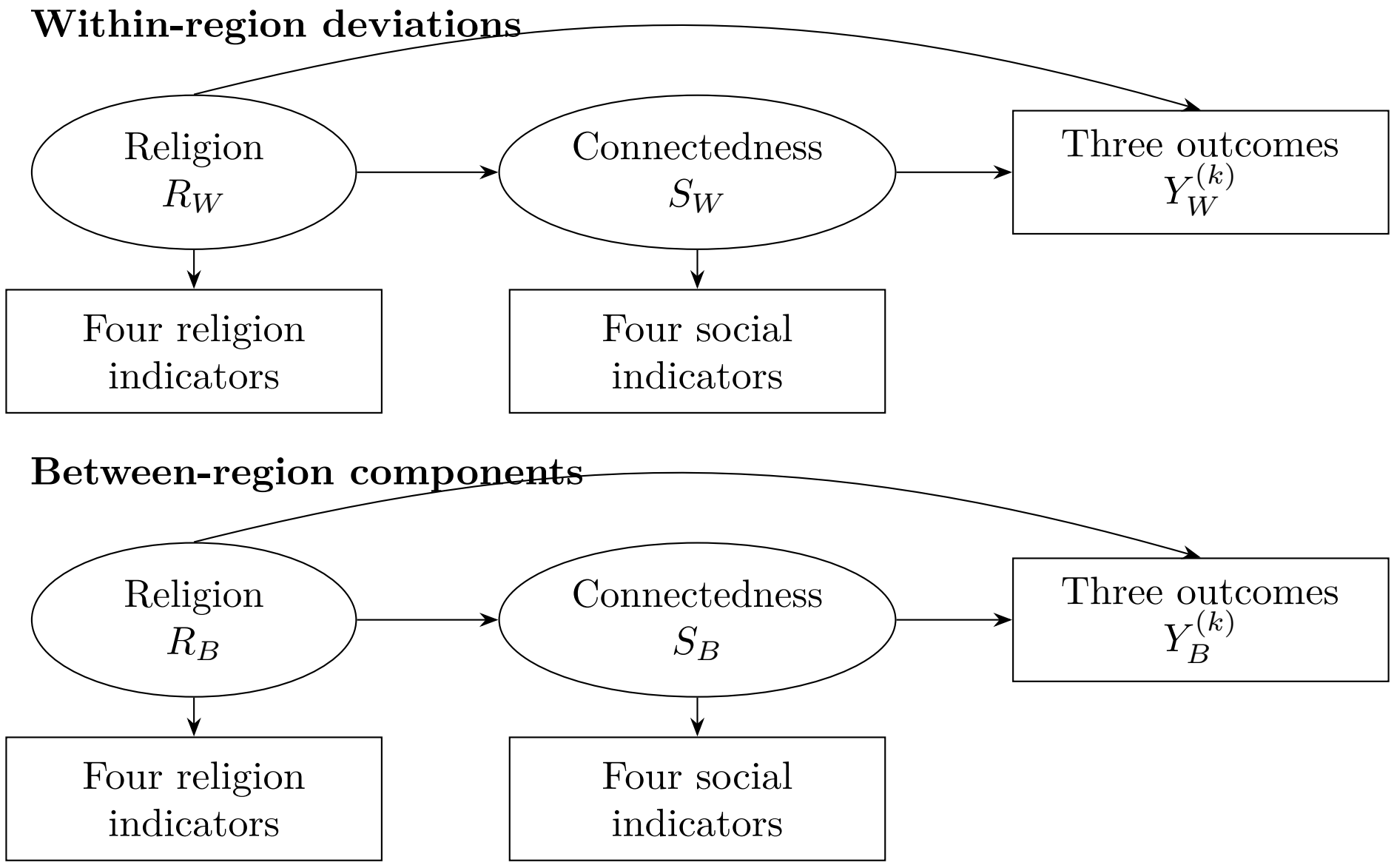
\begin{figure}[htbp]
\centering
\begin{tikzpicture}[>=Stealth,
 latent/.style={draw,ellipse,align=center,font=\small,minimum width=29mm,minimum height=12mm},
 observed/.style={draw,align=center,font=\small,text width=31mm,minimum height=11mm}]
\foreach \lev/\yy/\heading in {W/0/Within-region deviations,B/-4/Between-region components}{
  \node[font=\bfseries\small,anchor=west] at (-1.6,\yy+1.3) {\heading};
  \node[latent] (r\lev) at (0,\yy) {Religion\\$R_{\lev}$};
  \node[latent] (s\lev) at (4.5,\yy) {Connectedness\\$S_{\lev}$};
  \node[observed] (y\lev) at (9,\yy) {Three outcomes\\$Y^{(k)}_{\lev}$};
  \node[observed] (ri\lev) at (0,\yy-1.6) {Four religion\\indicators};
  \node[observed] (si\lev) at (4.5,\yy-1.6) {Four social\\indicators};
  \draw[->] (r\lev)--(s\lev);
  \draw[->] (s\lev)--(y\lev);
  \draw[->] (r\lev.north) to[bend left=15] (y\lev.north);
  \draw[->] (r\lev)--(ri\lev);
  \draw[->] (s\lev)--(si\lev);
}
\end{tikzpicture}
\caption{Attempted latent multilevel specification. Each grouped indicator
box represents four separate observed variables with their own loadings
and residuals; each outcome also has its own residual. No estimated
coefficients are shown.}
\label{fig:latent-specification}
\end{figure}

The unconstrained model converged but had three negative residual
variances: \texttt{believegod} at both levels and future \emph{happiness} at
the between-region level. Its comparative fit index (CFI) was 0.899 and root mean square error of approximation (RMSEA) was 0.124. Placing
zero lower bounds on every observed residual variance at both levels
removed negative estimates, but the same three variances were at the
boundary. Fit remained weak, with CFI 0.891 and RMSEA 0.129
(Table~\ref{tab:latent}). Neither model supports reliable interpretation
of its path coefficients.

\begin{table}[htbp]
\centering\small
\caption{Latent multilevel SEM diagnostics.}
\label{tab:latent}
\setlength{\tabcolsep}{5pt}
\renewcommand{\arraystretch}{1.25}
\begin{tabular}{@{}lrrrr@{}}
\toprule
Specification & Negative variances & At zero & CFI & RMSEA \\
\midrule
Unconstrained & 3 & 0 & 0.899 & 0.124 \\
Nonnegative variances & 0 & 3 & 0.891 & 0.129 \\
\bottomrule
\end{tabular}
\tabnote{Both fits converged on 5,122 region-years in 847 regions, with 74 degrees of freedom. Negative residual variances invalidate the unconstrained fit; boundary estimates and weak fit prevent interpretation of the bounded model's coefficients.}
\end{table}

\FloatBarrier

\end{document}